\documentclass[a4paper, 11pt, abstracton, notitlepage, oneside]{scrartcl}
\makeatletter

\usepackage{lmodern}

\usepackage[T1]{fontenc}
\usepackage[utf8]{inputenc}
\usepackage[onehalfspacing]{setspace}
\usepackage[headsepline]{scrlayer-scrpage}
\clearscrheadfoot 
\ohead{\pagemark}
\usepackage{layout}
\usepackage[english]{babel}
\usepackage[left=2cm,right=2cm,top=2.5cm,bottom=2.5cm]{geometry}

\usepackage[boxed, linesnumbered,noend]{algorithm2e}
\usepackage{adjustbox}
\usepackage{eurosym}
\usepackage{lscape}
\usepackage{nicefrac}
\usepackage{array}
\usepackage{paralist}
\usepackage{authblk}
\usepackage{graphicx}
\usepackage{booktabs}
\usepackage{placeins}
\usepackage{amsmath}
\usepackage{mathtools}
\usepackage{bm}
\usepackage{amssymb}
\usepackage{color}
\usepackage[acronym]{glossaries}
\usepackage{pgfplots}
\usepackage[normal]{threeparttable}
\usepackage{ulem}
\usepackage{natbib}
\usepackage[titletoc,title]{appendix}
\usepackage[hyphens]{url} 
\usepackage{ellipsis}
\usepackage{breqn} 
\usepackage{chngcntr}  
\usepackage{enumerate}
\usepackage{theorem}
\usepackage{etoolbox}
\usepackage{multirow}
\usepackage{colortbl}
\usepackage{subcaption}
\usepackage{rotating}
\usepackage{tikz}
\usetikzlibrary{arrows.meta}
\usetikzlibrary{backgrounds}
\usepackage{pgfplotstable}

\makeglossaries

\newcommand{\summe}[1]{\sum_{\scriptstyle\mathclap{#1}}}

\begin{document}
\newacronym{abk:aseag}{ASEAG}{'Aachener Straßenbahn und Energieversorgungs-AG'}
\newacronym[firstplural=battery electric vehicles (BEVs)]{abk:bev}{BEV}{battery electric vehicle}
\newacronym[firstplural=battery electric buses (BEBs)]{abk:beb}{BEB}{battery electric bus}
\newacronym{abk:bss}{BSS}{battery swap station}
\newacronym{abk:co2}{$\text{CO}_\text{2}$}{carbon dioxide}
%
\newacronym[firstplural=night charging buses (NCBs)]{abk:beb-on}{NCB}{night charging bus}
\newacronym[firstplural=night charging facilities (NCFs)]{abk:srf}{NCF}{night charging facility}
\newacronym{abk:evrptw}{EVRP-TW}{electric vehicle routing problem with time windows}
\newacronym{abk:evrp}{EVRP}{electric vehicle routing problem}
\newacronym{abk:eu}{EU}{European Union}
\newacronym{abk:evsp}{EVSP}{electric vehicle scheduling problem}
\newacronym{abk:elrp}{ELRP}{electric location routing problem}
\newacronym{abk:frlp}{FRLP}{flow refueling location problem}
\newacronym{abk:frlp-lfc}{FRLP-LFC}{Flow-Refueling-Location-Problem with Load-Flow-Control}
\newacronym[firstplural=opportunity charging buses (OCBs)]{abk:beb-oc}{OCB}{opportunity charging bus}
\newacronym{abk:ghg}{GHG}{greenhouse gas}
\newacronym{abk:hpc}{HPC}{High-Power-Charging}
\newacronym[firstplural=internal combustion engine buses (ICEBs)]{abk:iceb}{ICEB}{internal combustion engine bus}
%
%
%
\newacronym{abk:mdvsp}{MDVSP}{multi-depot vehicle scheduling problem}
\newacronym{abk:mcfp}{MCFP}{minimum-cost flow problem}
\newacronym{abk:mip}{MIP}{mixed integer problem}
\newacronym{abk:nox}{$\text{NO}_\text{x}$}{nitrogen oxide}
\newacronym{abk:no2}{$\text{NO}_\text{2}$}{nitrogen dioxide}
\newacronym{abk:od}{OD-pair}{origin and destination pair}
\newacronym{abk:pm}{$\text{PM}$}{particulate matter}
\newacronym{abk:pm2.5}{$\text{PM}_{2.5}$}{particulate matter ($2.5\,\mu m$)}
\newacronym{abk:pm10}{$\text{PM}_{10}$}{particulate matter ($10\,\mu m$)}
%
\newacronym[firstplural=charging facilities (CFs)]{abk:rf}{CF}{charging facility}
\newacronym{abk:soc}{SOC}{State-of-Charge}
\newacronym{abk:sdvsp}{SDVSP}{single-depot vehicle scheduling problem}
\newacronym{abk:tco}{TCO}{total cost of ownership}
\newacronym[firstplural=opportunity charging facilities (OCFs)]{abk:frf}{OCF}{opportunity charging facility}
\newacronym{abk:fs}{TS}{first or finale station of a bus line}
%
\newacronym[firstplural=vehicle scheduling problems (VSPs)]{abk:vsp}{VSP}{vehicle scheduling problem}
\newacronym{abk:vsplpr}{VSPLPR}{vehicle scheduling problem with length of path considerations}
%
%
%


\definecolor{rwth}   {RGB}{  0  84 159}
\definecolor{rwth-75}{RGB}{ 64 127 183}
\definecolor{rwth-50}{RGB}{142 186 229}
\definecolor{rwth-25}{RGB}{199 221 242}
\definecolor{rwth-10}{RGB}{232 241 250}

\definecolor{black}   {RGB}{  0   0   0}
\definecolor{black-75}{RGB}{100 101 103}
\definecolor{black-50}{RGB}{156 158 159}
\definecolor{black-25}{RGB}{207 209 210}
\definecolor{black-10}{RGB}{236 237 237}

\definecolor{magenta}   {RGB}{227   0 102}
\definecolor{magenta-75}{RGB}{233  96 136}
\definecolor{magenta-50}{RGB}{241 158 177}
\definecolor{magenta-25}{RGB}{249 210 218}
\definecolor{magenta-10}{RGB}{253 238 240}

\definecolor{yellow}   {RGB}{255 237   0}
\definecolor{yellow-75}{RGB}{255 240  85}
\definecolor{yellow-50}{RGB}{255 245 155}
\definecolor{yellow-25}{RGB}{255 250 209}
\definecolor{yellow-10}{RGB}{255 253 238}

\definecolor{petrol}   {RGB}{  0  97 101}
\definecolor{petrol-75}{RGB}{ 45 127 131}
\definecolor{petrol-50}{RGB}{125 164 167}
\definecolor{petrol-25}{RGB}{191 208 209}
\definecolor{petrol-10}{RGB}{230 236 236}

\definecolor{turkis}   {RGB}{  0 152 161}
\definecolor{turkis-75}{RGB}{  0 177 183}
\definecolor{turkis-50}{RGB}{137 204 207}
\definecolor{turkis-25}{RGB}{202 231 231}
\definecolor{turkis-10}{RGB}{235 246 246}

\definecolor{grun}   {RGB}{ 87 171  39}
\definecolor{grun-75}{RGB}{141 192  96}
\definecolor{grun-50}{RGB}{184 214 152}
\definecolor{grun-25}{RGB}{221 235 206}
\definecolor{grun-10}{RGB}{242 247 236}

\definecolor{maigrun}   {RGB}{189 205   0}
\definecolor{maigrun-75}{RGB}{208 217  92}
\definecolor{maigrun-50}{RGB}{224 230 154}
\definecolor{maigrun-25}{RGB}{240 243 208}
\definecolor{maigrun-10}{RGB}{249 250 237}

\definecolor{orange}   {RGB}{246 168   0}
\definecolor{orange-75}{RGB}{250 190  80}
\definecolor{orange-50}{RGB}{253 212 143}
\definecolor{orange-25}{RGB}{254 234 201}
\definecolor{orange-10}{RGB}{255 247 234}

\definecolor{rot}   {RGB}{204   7  30}
\definecolor{rot-75}{RGB}{216  92  65}
\definecolor{rot-50}{RGB}{230 150 121}
\definecolor{rot-25}{RGB}{243 205 187}
\definecolor{rot-10}{RGB}{250 235 227}

\definecolor{bordeaux}   {RGB}{161  16  53}
\definecolor{bordeaux-75}{RGB}{182  82  86}
\definecolor{bordeaux-50}{RGB}{205 139 135}
\definecolor{bordeaux-25}{RGB}{229 197 192}
\definecolor{bordeaux-10}{RGB}{245 232 229}

\definecolor{violett}   {RGB}{ 97  33  88}
\definecolor{violett-75}{RGB}{131  78 117}
\definecolor{violett-50}{RGB}{168 133 158}
\definecolor{violett-25}{RGB}{210 192 205}
\definecolor{violett-10}{RGB}{237 229 234}

\definecolor{lila}   {RGB}{122 111 172}
\definecolor{lila-75}{RGB}{155 145 193}
\definecolor{lila-50}{RGB}{188 181 215}
\definecolor{lila-25}{RGB}{222 218 235}
\definecolor{lila-10}{RGB}{242 240 247}


\title{\large On the Integration of Battery Electric Buses into
	Urban Bus Networks}

\author[1]{\normalsize Nicolas Dirks\footnote{corresponding author}}
\author[2]{\normalsize Maximilian Schiffer}
\author[1]{\normalsize Grit Walther}

	\small 
	\affil[1]{Chair of Operations Management, School of Business and Economics, RWTH Aachen University, 52072 Aachen, Germany}
	\affil[2]{TUM School of Management, Technical University of Munich, 80333 Munich, Germany}

	\affil[ ]{
	\scriptsize
	nicolas.dirks@om.rwth-aachen.de,
	schiffer@tum.de,
	walther@om.rwth-aachen.de}

\date{}


\maketitle
\begin{abstract}
\begin{singlespace}
{\small\noindent 
	
	Cities all around the world struggle with urban air quality due to transportation related emissions. In public transport networks, replacing internal combustion engine buses by electric buses provides an opportunity to improve air quality. Hence, many bus network operators currently ask for an optimal transformation plan to integrate battery electric buses into their fleet. Ideally, this plan also considers the installation of necessary charging infrastructure to ensure a fleet’s operational feasibility.
	%
	Against this background, we introduce an integrated modeling approach to determine a cost-optimal, long-term, multi-period transformation plan for integrating battery electric buses into urban bus networks. Our model connects central strategic and operational decisions. We minimize total cost of ownership and analyze potential reductions of nitrogen oxide emissions. 
	%
	Our results base on a case study of a real-world bus network and show that a comprehensive integration of battery electric buses is feasible and economically beneficial. By analyzing the impact of battery capacities and charging power on the optimal fleet transformation, we show that medium-power charging facilities combined with medium-capacity batteries are superior to networks with low-power or high-power charging facilities.

\smallskip}
\end{singlespace}
\end{abstract}
\newpage
\section{Introduction}
Cities around the world struggle with low air quality, which is to a significant extend caused by transportation related emissions. In this course, the European Union issued inner-city emission thresholds to achieve a reduction of air pollutant concentrations~(cf. \citealt{EC2008}), which requires to curb transportation related emissions. Here, the utilization of \glspl{abk:bev} that cause zero local emissions remains a viable option to meet these thresholds. Accordingly, for public transport, \glspl{abk:beb} provide a promising alternative to \glspl{abk:iceb} with potential reductions of \gls{abk:nox} emissions and savings in operational~(fuel) costs. Known challenges of electric vehicles such as limited driving ranges and needs for recharging appear to be manageable for \glspl{abk:beb} as their routes are predetermined and thus, fleet operators are able to consider these limitations a-priori. Accordingly, many public transport authorities aim at an integration of \glspl{abk:beb} into their public bus fleet. Amongst others, the cities of Paris and Nottingham consider and test the partial~(Paris) or full~(Nottingham) electrification of their public bus network~(see, e.g., \citealt{Zeus2017}).

In this course, a transformation towards electrified bus fleets is a central task for many fleet operators resulting in the following planning tasks: At strategic level, operators must decide on the transformation of the bus fleet and the investment into sufficient charging infrastructure. Transforming the bus fleet over time requires strategic decisions on purchases and sales of buses linked to decisions on drivetrain technologies, battery capacities, and charging concepts. Installing charging infrastructure requires decisions on the timing of the installation as well as on the locations and characteristics of charging facilities. At operational level, operators need to ensure operational feasibility of the bus timetable, i.e., of service trips specified by bus lines and service times. To guarantee operational feasibility, the right buses need to get assigned to these service trips, while considering limited driving ranges of \glspl{abk:beb}, and~(partial) recharging operations. Accordingly, integrating \glspl{abk:beb} into urban bus networks requires a transformation plan that links strategic bus network design decisions with operational feasibility constraints.

Against this background, we introduce an integrated modeling approach for a cost-optimal, long-term, multi-period transformation plan to integrate \glspl{abk:beb} into urban bus networks. We minimize a fleet transformation's \gls{abk:tco} based on a \gls{abk:mip} that considers strategic and operational decisions simultaneously: At strategic level, we provide a transformation plan with decisions on purchases and sales of buses, specified by drivetrain technologies, battery capacities, charging concepts, and decisions on charging infrastructure network design. At operational level, this transformation plan accounts for the assignment of buses to trip sequences of a vehicle schedule, while considering (partial) recharging operations to ensure operational feasibility. Our modeling approach constitutes a generic tool that can readily be applied to different bus networks by academics or practitioners as it remains solvable with of the shelf optimization software.

 We apply our model to the real-world bus network of the city of Aachen, Germany. Our results allow to analyze the resulting transformation strategy as well as to conduct sensitivity analyses for design-critical parameters, e.g., battery capacities and low- or high-charging power concepts. Moreover, we assess reductions of \gls{abk:nox} emissions achieved by an integration of \glspl{abk:beb}. Based on these results, we provide several managerial insights that point towards promising directions for future \gls{abk:beb} applications.

The remainder of this paper is structured as follows. In Section~\ref{sect:literature}, we give an overview of related literature, before we develop our methodology in Section~\ref{sect:model}. Section~\ref{sect:caseStudy} describes our case study and experimental design. Section~\ref{sect:results} discusses our results, and Section~\ref{sect:conclusion} concludes this paper.
\section{Literature Review}\label{sect:literature}
Our work relates to the broad field of electric vehicle routing and scheduling, and charging network design. To keep this paper concise, we point the interested reader to the recent surveys of \citet{SchifferSchneiderEtAl2018b} and \cite{Shen2019}, and focus only on \gls{abk:beb} specific work in the following.
\paragraph{\gls{abk:tco} analysis} \gls{abk:tco} analyses have widely been used for descriptive analyses of the competitiveness of electric vehicles~(see, e.g., \citealt{FengFigliozzi2013}), incorporating different aspects such as investment, energy, operational and maintenance cost; herein approximating vehicle operations and strategic decisions. In this field, only a few studies focused on \gls{abk:tco} analysis for \glspl{abk:beb}.
\citet{Goehlich2014} focused on technology assessment for \glspl{abk:beb}, while \citet{Pihaltie2014} proposed a \gls{abk:tco} model including a component, vehicle, and traffic system analysis. Both studies found that the lowest \glspl{abk:tco} are realized for buses with low-capacity batteries combined with high-power charging facilities.  In contrast, \citet{Nurhadi2014} identified low-power charging facilities combined with high-capacity battery buses as the most beneficial combination using a similar methodology. \citet{Lajunen2018} compared different charging concepts and found that opportunity charging buses show lower \gls{abk:tco} than overnight charging buses.

Concluding, some \gls{abk:tco} analyses exist for \glspl{abk:beb} but lack the explicit consideration of strategic network design decisions and operational constraints. Hence, their results depend heavily on the chosen approximations and parameters and may lead to conflicting outcomes~(see, e.g., \citealt{Pihaltie2014}, \citealt{Nurhadi2014}).
\paragraph{Strategic planning approaches} Strategic planning approaches focus on cost-optimal decisions for bus fleet composition or charging infrastructure planning. \citet{Kunith2017} focused on locating charging facilities for an entire bus network, taking a decision for each bus line separately, while including decisions on the battery power of buses and the charging power of charging facilities. In a similar approach, \citet{Xylia2017} accounted for different drivetrain technologies, assigning one specific type to each bus line. \citet{Li2018a} focused on determining an optimal bus fleet composition, including partial recharging considerations. \citet{Wei2018} focused on finding an optimal fleet composition including the installation of charging facilities for a single point in time, while considering a given vehicle schedule with dead-heading. \citet{Islam2019a} minimized costs for vehicle investments, depot charging facilities, and external emissions for a fleet replacement process in which only a limited amount of vehicles can be replaced with \glspl{abk:iceb}. \citet{Pelletier2019} proposed a multi-period fleet transformation model that minimizes total costs for depot and fast charging buses. They approximated daily operations for a given vehicle schedule as well as costs for charging infrastructure. This approximation does not ensure operational feasibility as specific charging facility locations as well as energy balances of \glspl{abk:beb} along routes are neglected. \citet{Lin2019b} proposed a model for locating charging facility parks that consist of multiple charging facilities and considered interdependencies with the electricity grid. \citet{Li2019a} proposed a time-space-energy network to locate charging facilities while accounting for the bus fleet composition and external costs of emissions. \citet{An2020a} proposed a stochastic model to optimize charging facility locations and the bus fleet size while considering time-dependent electricity prices.  

Concluding, several strategic planning approaches exist, but none of these approaches considers all necessary planning components. Most approaches do not guarantee operational feasibility for the identified solutions. The approaches that consider operational feasibility regard only limited settings~(e.g., ignore dead-heading of buses) or do not account for a multi-period time horizon. The few approaches that account for strategic decisions and operational feasibility exist in the field of electric location-routing problems~(see, e.g., \citealt{SchifferWalther2017}) and cannot be applied to our planning problem as they neglect bus network specific characteristics. 
\paragraph{Operational planning approaches} Operational planning approaches resemble electric vehicle scheduling problems and aim to determine an optimal vehicle schedule that requires a minimum number of buses for a given network and charging infrastructure.
\citet{Li2014} focused on a single-depot vehicle scheduling problem with full recharging or battery swapping for \glspl{abk:beb}, while \citet{Adler2017} considered multiple depots but no consecutive recharging operations. \citet{Wen2016} considered multiple depots, consecutive recharging operations, as well as full and partial recharging. \citet{Niekerk2017} incorporated electricity prices, battery depreciation, and various recharging options. Recently, \citet{Tang2019a} analyzed robust scheduling strategies to account for stochastic traffic conditions, and \citet{Yao2020a} proposed a heuristic to minimize total costs of a bus network operated by depot charging buses.

Recently, first operational approaches started to integrate strategic decisions and focused on the integration of charging station network design decisions. \citet{Rogge2018} focused on determining a cost-minimal vehicle schedule while estimating the number of depot charging facilities that are necessary to recharge buses.  \citet{Liu2020a} proposed a bi-objective model to schedule buses and to install fast charging facilities at first or final stations of a bus line, herein accounting for both dead-heading and partial recharging.

Concluding, some operational planning approaches exist that determine or account for vehicle schedules. While some of them aim to integrate strategic decisions for the design of charging station infrastructure, none extends to integrate fleet transformation characteristics.
\paragraph{Conclusion} Table~\ref{tab:check} shows the relation between our work and existing approaches that remain focused on either strategic or operational aspects of the planning problem. As can be seen, strategic modeling approaches so far lack either a multi-period time horizon, operational feasibility, or charging station infrastructure design decisions. In particular, first approaches that strive towards an integrated modeling approach (\citealt{Wei2018}, \citealt{Li2019a}) lack a multi-period investment decision and partial recharging. Operational approaches lack among others strategic fleet transformation decisions, charging infrastructure design decisions, and often a total cost perspective. The only operational approach that strives towards an integrated model (\citealt{Liu2020a}) integrates charging infrastructure design decisions but still neglects all fleet transformation decisions as well as a total cost perspective.

With this work, we present a comprehensive and realistic planning model that covers all requirements as listed in Table~\ref{tab:check}. Specifically, we: \textit{i)} perform a comprehensive \gls{abk:tco} optimization, considering all decision relevant strategic costs as well as anticipated operational costs for using \glspl{abk:beb}; \textit{ii)} develop an integrated approach that considers strategic and operational decisions simultaneously; \textit{iii)}~derive a transformation plan to integrate \glspl{abk:beb} in a heterogeneous bus fleet by regarding purchases and sales of buses with alternative drivetrain technologies, battery capacities, and charging concepts; \textit{iv)} optimize the time-dependent installation of charging infrastructure; and \textit{v)} ensure operational feasibility by considering a given vehicle schedule of a bus timetable including (partial) recharging decisions for \glspl{abk:beb}. We apply this model to a real-world case study and present managerial insights, including the  reductions of \gls{abk:nox} emissions that result from the integration of \glspl{abk:beb}.

\begin{table*}[ht!]
	\centering
	\begin{threeparttable}
		\centering
		\newcommand{\PreserveBackslash}[1]{\let\temp=\\#1\let\\=\temp}
		\newcolumntype{C}[1]{>{\PreserveBackslash\centering}p{#1}}
		\setlength{\abovecaptionskip}{0.5ex}
		\caption{Overview of related modeling approaches.}
		\label{tab:check}
		\scriptsize
		\def\c{\checkmark}
		\addtolength{\tabcolsep}{-3pt}
		\singlespacing
		\begin{tabular}{rcccccccccccccccccc}
			\toprule
			& \multicolumn{9}{c}{strategic} & \multicolumn{8}{c}{operational} & \\		
			\cmidrule(lr{0.6em}){2-10}\cmidrule(lr{0.6em}){11-18}				
			&\multicolumn{1}{l}{\begin{turn}{82}\citealt{Kunith2017}\end{turn}} &\multicolumn{1}{l}{\begin{turn}{82}\citealt{Xylia2017}\end{turn}} &\multicolumn{1}{l}{\begin{turn}{82}\citealt{Li2018a}\end{turn}} &\multicolumn{1}{l}{\begin{turn}{82}\citealt{Islam2019a}\end{turn}} &\multicolumn{1}{l}{\begin{turn}{82}\citealt{Pelletier2019}\end{turn}} &\multicolumn{1}{l}{\begin{turn}{82}\citealt{Lin2019b}\end{turn}} &\multicolumn{1}{l}{\begin{turn}{82}\citealt{An2020a}\end{turn}} &\multicolumn{1}{l}{\begin{turn}{82}\citealt{Wei2018}\end{turn}} &\multicolumn{1}{l}{\begin{turn}{82}\citealt{Li2019a}\end{turn}} &\multicolumn{1}{l}{\begin{turn}{82}\citealt{Li2014}\end{turn}} &\multicolumn{1}{l}{\begin{turn}{82}\citealt{Wen2016}\end{turn}} &\multicolumn{1}{l}{\begin{turn}{82}\citealt{Adler2017}\end{turn}} &\multicolumn{1}{l}{\begin{turn}{82}\citealt{Niekerk2017}\end{turn}} &\multicolumn{1}{l}{\begin{turn}{82}\citealt{Tang2019a}\end{turn}} &\multicolumn{1}{l}{\begin{turn}{82}\citealt{Yao2020a}\end{turn}} &\multicolumn{1}{l}{\begin{turn}{82}\citealt{Rogge2018}\end{turn}} &\multicolumn{1}{l}{\begin{turn}{82}\citealt{Liu2020a}\end{turn}} &\multicolumn{1}{l}{\begin{turn}{82}this paper\end{turn}} \\
			\midrule	
			total costs		 	    		&\c&\c&\c&\c&\c&\c&\c&\c&\c&- &- &- &- &- &\c&\c&- &\c \\
			multi-period					&- &- &\c&\c&\c&\c&- &- &- &- &- &- &- &- &- &- &- &\c \\
			heterogeneous bus fleet			&- &\c&\c&\c&\c&- &\c &\c&\c&- &\c&\c&\c&- &\c&\c&- &\c \\			
			bus purchases and sales 		&- &- &\c&\c&\c&- &- &- &- &- &- &- &- &- &- &- &- &\c \\				
			siting charging facilities		&\c&\c&- &- &- &\c&\c&\c&\c&- &- &- &- &- &- &- &\c&\c \\
			operational feasibility			&\c&\c&\c&- &- &- &- &\c&\c&\c&\c&\c&\c&\c&\c&\c&\c&\c \\
			partial recharging				&\c&\c&\c&- &- &- &\c&- &- &- &\c&- &\c&- &- &- &\c&\c \\				
			dead-heading					&- &- &- &- &- &- &- &\c&\c&\c&\c&\c&\c&\c&\c&\c&\c&\c \\
			\bottomrule		
		\end{tabular}				
		\centering
	\end{threeparttable}
\end{table*}
\section{Methodology}\label{sect:model}
In this section, we introduce our methodology to determine an optimal transformation plan for integrating \glspl{abk:beb} into urban bus networks. We first present our problem setting in Section~\ref{sect:model-assumptions}, before we develop a corresponding mixed-integer linear model in Section~\ref{sect:model-formulation}.
\subsection{Problem setting}\label{sect:model-assumptions}
We focus on the transformation of a bus fleet, which is used to operate a public urban bus network from the point of view of its fleet operator. Specifically, we decide on the sale and purchase of buses over time with respect to the bus type, its drivetrain, and in case of \glspl{abk:beb} its battery capacity. Additionally, we decide on investments into necessary charging infrastructure to operate \glspl{abk:beb}. Here, we decide on the required charging infrastructure, the location of charging facilities and on the used charging concept, i.e., if buses are only charged at the depot (overnight charging) or if partial recharging at bus stations (opportunity charging) is allowed. Moreover, we consider the operational requirements of the bus network and guarantee that the fleet is at any time able to operate according to the given bus timetable.

In this setting, we consider a heterogeneous bus fleet with different drivetrain technologies, namely \glspl{abk:iceb} and \glspl{abk:beb} with a homogeneous economic vehicle lifetime. We focus on a finite planning horizon during which the operator may sell \glspl{abk:iceb} to replace them with buses of the same technology or with a \gls{abk:beb}. Once an operator decides to replace an \gls{abk:iceb} by a \gls{abk:beb}, the \gls{abk:beb} remains in the fleet until the end of its economic lifetime.  We assume that the lifetimes of the \glspl{abk:beb}' batteries are shorter than the economic lifetime of the \glspl{abk:beb}. Thus, the operator has to account for battery replacements. To account for beginning and end of horizon effects in the planning time horizon, we account for an initial purchase of the initial bus fleet's value as well as for the salvage value of the bus fleet at the end of the planning horizon depending on the ages of the buses. To calculate discounted cash flows we use a linear depreciation rate.

While \glspl{abk:iceb} do not require refueling during daily operations, \glspl{abk:beb} may need to recharge while operating. Herein, the necessity to recharge depends on the bus line, the driving ranges, and the battery capacities of buses, respectively. Accordingly, we account for decisions on battery capacities.

We decide on the number and locations of two different types of charging facilities. While \glspl{abk:srf} exclusively get installed at the central depot, we assume that the high-power, fast-charging \glspl{abk:frf} (e.g., pantographs) can be sited at initial and final stations of bus lines. Once installed, charging facilities cannot get uninstalled.

Accordingly, we differentiate buses into two different types, \glspl{abk:beb-on} and \glspl{abk:beb-oc}. While \glspl{abk:beb-on} exclusively recharge overnight at the depot, \glspl{abk:beb-oc} can additionally recharge during the day at \glspl{abk:frf} during~(short) dwell-times in between operations. A distinction between \glspl{abk:beb-on} and \glspl{abk:beb-oc} is necessary because \glspl{abk:beb-oc} must be equipped with special batteries and on-board charging devices to be able to recharge at \glspl{abk:frf}

In order to consider operational requirements, we embed a vehicle schedule that is feasible for \glspl{abk:iceb} based on a given bus timetable. The bus timetable includes several service trips that must be operated. A service trip represents a driving activity with passengers on-board starting at an initial and ending at a final station of a bus line. Accordingly, a service trip is defined by start and end times at the respective initial and final bus stations.

\begin{figure}[!ht]
	\centering
	\begin{minipage}{\textwidth}
		\centering
		\scalebox{.85}{
			\begin{tikzpicture}[scale=0.65,framed]
	\draw[-Latex,very thick] (-1.00,1.00) -- (-1.00,-2.00);
	\node[rotate=90] (y) at (-1.25,-0.25){space};
	\draw[-Latex,very thick] (-1.03,1.00) -- (2.00,1.00);
	\node (y) at (0.25,1.25){time};
	\node[rectangle,fill] (0) at (0.00,0.00){};
	\node[circle,fill,scale=0.75] (1) at (1.00,-4.00){};
	\draw[dashed,-Latex] (0) edge (1);
	\node[circle,draw,scale=0.75] (2) at (1.25,-5.00){};
	\draw[-Latex] (1) edge (2);
	\node[circle,draw,scale=0.75] (3) at (1.50,-6.00){};
	\draw[-Latex] (2) edge (3);
	\node[circle,draw,scale=0.75] (4) at (1.75,-7.00){};
	\draw[-Latex] (3) edge (4);
	\node[circle,fill,scale=0.75] (5) at (2.00,-8.00){};
	\draw[-Latex] (4) edge (5);
	\node[circle,fill,scale=0.75] (6) at (3.00,-8.00){};
	\draw[dotted,-Latex] (5) -- (6);
	\node[circle,draw,scale=0.75] (7) at (3.25,-7.00){};
	\draw[-Latex] (6) edge (7);
	\node[circle,draw,scale=0.75] (8) at (3.50,-6.00){};
	\draw[-Latex] (7) edge (8);
	\node[circle,draw,scale=0.75] (9) at (3.75,-5.00){};
	\draw[-Latex] (8) edge (9);
	\node[circle,fill,scale=0.75] (10) at (4.00,-4.00){};
	\draw[-Latex] (9) edge (10);
	\node[circle,fill,scale=0.75] (11) at (4.50,-2.00){};
	\draw[dashed,-Latex] (10) edge (11);
	\node[circle,fill,scale=0.75] (12) at (6.50,-2.00){};
	\draw[dotted,-Latex] (11) edge (12);
	\node[circle,draw,scale=0.75] (13) at (6.75,-3.00){};
	\draw[-Latex] (12) edge (13);
	\node[circle,draw,scale=0.75] (14) at (7.00,-4.00){};
	\draw[-Latex] (13) edge (14);
	\node[circle,fill,scale=0.75] (15) at (7.25,-5.00){};
	\draw[-Latex] (14) edge (15);
	\node[circle,fill,scale=0.75] (16) at (9.25,-5.00){};
	\draw[dotted,-Latex] (15) -- (16);
	\node[circle,draw,scale=0.75] (17) at (9.50,-4.00){};
	\draw[-Latex] (16) edge (17);
	\node[circle,draw,scale=0.75] (18) at (9.75,-3.00){};
	\draw[-Latex] (17) edge (18);
	\node[circle,fill,scale=0.75] (19) at (10.00,-2.00){};
	\draw[-Latex] (18) edge (19);
	\node[rectangle,fill] (20) at (10.50,0.00){};
	\draw[dashed,-Latex] (19) edge (20);
	\node[rectangle,fill,label=east:\footnotesize{depot}] (a) at (5.50,-6.00){};
	\node[circle,fill,scale=0.75,label=east:\footnotesize{first / final station}] (b) at (5.50,-6.5){};
	\node[circle,draw,scale=0.75,label=east:\footnotesize{bus stop}] (c) at (5.50,-7.0){};
	\draw[-Latex] (5.25,-7.5) -- (5.75,-7.5);
	\node[label=east:\footnotesize{passenger transport}] (d) at (5.50,-7.5){};
	\draw[dotted,-Latex] (5.25,-8.0) -- (5.75,-8.0);	
	\node[label=east:\footnotesize{dwell-time}] (d) at (5.50,-8.0){};
	\draw[dashed,-Latex] (5.25,-8.5) -- (5.75,-8.5);
	\node[label=east:\footnotesize{dead-heading}] (d) at (5.50,-8.5){};
	\draw[draw] (5.00,-5.50) edge (5.00,-8.75);
	\draw[draw] (5.00,-5.50) edge (10.50,-5.50);
	\draw[draw] (5.00,-8.75) edge (10.50,-8.75);
	\draw[draw] (10.50,-5.50) edge (10.50,-8.75);
\end{tikzpicture}
		}
	\end{minipage}
	\caption{Exemplary trip sequence of a vehicle schedule.}
	\label{fig:sequence}
\end{figure}
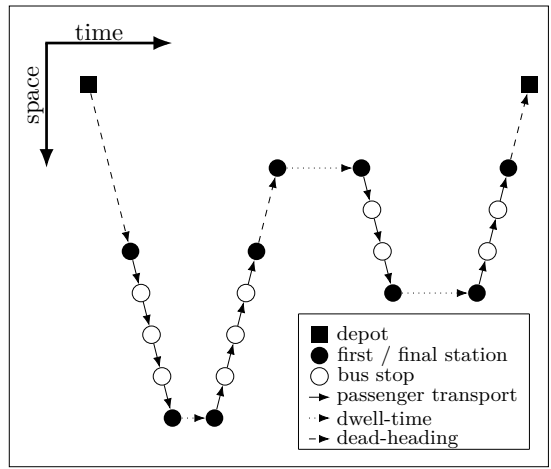

We use the bus timetable to create a vehicle schedule (cf. Appendix~\ref{apx:A}) that consists of several trip sequences. A trip sequence denotes a specific order of service trip operations, dead-heading operations, and dwell-time operations that are performed by a single bus during one day. Herein, dead-heading operations constitute driving activities without passengers, which are used to transfer the bus from the final station of a trip to the initial station of the next trip. During dwell-time operations, buses pause at the first station of a trip in order to stay in line with the bus timetable. If such a station is equipped with a charging facility, an \gls{abk:beb-oc} may use available dwell-time to recharge its battery before departing to its next operation. Figure~\ref{fig:sequence} 
shows an example of one trip sequence containing four trips. Each trip sequence is operated by a dedicated bus, such that the number of trip sequences of the vehicle schedule indicates the number of buses required to operate the bus network. A bus starts and ends operating a specific trip sequence at the depot. We assume that any \gls{abk:beb} starts with a full battery when leaving the depot in the morning, and model the amount of energy charged during a recharging process linear to the available recharging time.

As we aim at a conservative assessment of \glspl{abk:beb}, we neglect potential additional revenue due to power grid services such as demand side management, vehicle to grid operations, and participation in the energy balancing market, as well as additional revenue due to a second life usage of batteries. We further neglect insurance and taxes as they typically do not differ between \glspl{abk:iceb} and \glspl{abk:beb}. Since the total number of buses and bus drivers is predetermined by the vehicle schedule and remains equal independent of the bus fleet composition, we do not account for labor costs.

Within this setting, we seek for a cost optimal transformation plan that indicates i) which buses should be replaced in which time period and with which bus type, ii) at which locations charging stations should be built in which time step, and iii) which bus should operate which trip sequence in which time step, such that iv) all bus to trip sequence allocations remain feasible over the planning horizon. 

A few comments on this problem setting are in order. First, we assume the lifetime of \glspl{abk:beb} and \glspl{abk:iceb} to be equal. As \glspl{abk:beb} are expected to have a higher lifetime than \glspl{abk:iceb} due to less wearing parts~(see, e.g., \citealt{DavisFigliozzi2013}), this assumption is in favor for \glspl{abk:iceb} and represents a conservative estimate for \glspl{abk:beb}. 
Second, we assume that we can always sell an \gls{abk:iceb}, while a \gls{abk:beb} remains in the fleet until the end of its lifetime once it is purchased. While this assumption clearly limits the decision space of the operator, it is in line with the scope of our study, which aims at analyzing the transformation towards an electrified bus fleet within a finite planning horizon. Third, we assume that charging facilities do not get uninstalled once they have been built. For charging facilities installed at the depot, no alternative locations exist, such that the assumption is not limiting in this case. For charging facilities at bus stations, the assumption is in line with current practice, where investments into charging facilities are treated as irrevocable strategic decisions, in particular since the placement of fast charging facilities requires significant additional investments for high voltage installations. Fourth, we assume that a bus starts with a fully charged battery in the morning. This does not affect the applicability of our results in practice as the buses usually have a sufficiently long break during the night such that they can be fully recharged. Fifth, we embed a predetermined vehicle schedule, which remains unchanged, in our model. This assumption is in line with transition processes in practice, where fleet operators may not be willing to take the effort to frequently modify their operation during the transformation period. Moreover, using a \gls{abk:iceb}-optimized schedule represents a conservative assumption for \glspl{abk:beb}, as a more cost-efficient assignment for \glspl{abk:beb} may exist. Sixth, we assume a linear recharging function. We fit this function such that it yields a worst-case estimate for the original recharging function, which shows a slightly concave shape.
\subsection{Model}\label{sect:model-formulation}
We now introduce a \gls{abk:mip} to formalize the problem setting outlined above. This \gls{abk:mip} accounts for discrete, time-dependent decisions over a finite time horizon~$\overline{\mathcal{T}} = \lbrace t_{0},...,t_{\text{n+1}} \rbrace$, where~$t_0$ and~$t_{n+1}$ denote synthetic accounting periods in which the purchase costs of the original fleet and the salvage value of the transformed fleet are considered, but no decisions are taken. $\mathcal{T} \subset \overline{\mathcal{T}}= \lbrace t_{1},...,t_{\text{n}} \rbrace$ denotes all time steps in which decisions can be taken. Without further notice, we index time-dependent variables and parameters by~$t$. In the following, we provide the notation of sets and decision variables. Table~\ref{tab:notation} summarizes our model parameters, sets, and decision variables. 

We introduce different sets, one for each bus type, to avoid a complex tracking of individual buses: let~$\mathcal{K}$ denote the set of all bus types, with $\mathcal{K}^{\text{ICEB}} \subseteq \mathcal{K}$ being the set of all \gls{abk:iceb} types and $\mathcal{K}^{\text{BEB}} \subseteq \mathcal{K}$ being the set of all \gls{abk:beb} types. 
\begin{table*}[!hb]
	\centering
	\begin{threeparttable}
		\centering
		\setlength{\abovecaptionskip}{0.5ex}
		\addtolength{\tabcolsep}{-4pt}		
		\caption{Notation of the model.}
		\label{tab:notation}
		\scriptsize
		\begin{tabular}{rll} 					
			\toprule
			\multicolumn{3}{l}{\textbf{Sets}}\\
			$\mathcal{K}$ & set of bus types & $k \in \mathcal{K}$\\
			$\mathcal{K}^{\text{ICEB}}$ & set of \gls{abk:iceb} types & $k \in \mathcal{K}^{\text{ICEB}}$\\
			$\mathcal{K}^{\text{BEB}}$ & set of \gls{abk:beb} types & $k \in \mathcal{K}^{\text{BEB}}$\\
			$\mathcal{K}^{\text{n}}$ & set of night charging \glspl{abk:beb} & $k \in \mathcal{K}^{\text{n}}$\\
			$\mathcal{K}^{\text{o}}$ & set of opportunity charging \glspl{abk:beb} & $k \in \mathcal{K}^{\text{o}}$\\
			$\mathcal{R}$  & set of potential charging facility locations & $i \in \mathcal{R}$\\
			$\mathcal{T}$ & set of time periods & $t \in \mathcal{T}$\\
			$\mathcal{S}$ & set of trip sequences & $s \in \mathcal{S}$\\
			$\mathcal{L}$ & set of labels of all trips & $l \in \mathcal{L}$\\
			$\mathcal{N}_{s}$ & set of labels $l\in\mathcal{L}$ of all trips that belong to trip sequence~$s$ & $i \in \mathcal{N}_{s}$\\
			$\mathcal{A}_{s}$ & set of tuples $(i,j)$ denoting that trip $j$ succeeds trip $i$ & $(i,j) \in \mathcal{A}_{s}$\\
			\multicolumn{3}{l}{\textbf{Decision variables}}\\
			$n_{kt}$ & integer: number of buses of type $k$ in fleet & - \\
			$p_{kt}$ & integer: number of buses of type $k$ purchased & - \\
			$a_{t}$ & integer: number of depot charging facilities & - \\
			$y_{it}$ & binary: indicates if a charging facility is available at location $i$ & - \\			
			$x_{skt}$ & binary: indicates if trip sequence $s$ is operated by a bus of type $k$ & - \\
			$\Theta_{st}$ & actual/fictional required battery capacity for operating trip sequence $s$ & kWh\\
			$q_{it}$ & battery state of charge before operating trip $i$ & kWh\\
			$w_{it}$ & amount of energy recharged before operating trip $i$ & kWh\\
			\multicolumn{3}{l}{\textbf{Parameters}}\\
			$c_{k}^{\text{b}}$ & costs for purchasing a bus of type $k$ & \euro\\
			$c_{kt}^{\text{q}}$ & costs for purchasing a battery per unit for bus type $k$ & \euro/kWh\\
			$C^{\text{fleet}}_{0}$ & value of the initial bus fleet & \euro\\
			$v^{\text{b}}$ & salvage revenue due to selling a bus of type $k$ & \euro\\
			$c^{\text{o}}$ & costs for installing a opportunity charging facility & \euro\\
			$c^{\text{n}}$ & costs for installing a night charging facility & \euro\\
			$c^{\text{om}}$ & annual costs for maintenance of an opportunity charging facility & \euro/a\\
			$c^{\text{nm}}$ & annual costs for maintenance of a night charging facility & \euro/a\\
			$\eta$ & annual days of operation & d \\
			$c_{k}^{\text{bm}}$ & costs for maintenance per distance covered by a bus of type $k$ & \euro/km\\
			$c_{k}^{\text{e}}$ & costs for energy per unit consumed by a bus of type $k$ & \euro/kWh\\
			$g$ & discount rate & - \\
			$e^{b}$ & economic lifetime of a bus & a\\
			$e^{q}$ & economic lifetime of a battery & a\\
			$h_{k}$ & holding period of a bus of type $k$ & a\\
			$Q_{k}$ & installed battery capacity in \gls{abk:beb} type $k$ & kWh\\
			$\mu$ & available percentage of battery capacity & - \\
			$r$ & charging power of a network charging facility & kW\\
			$\lambda_{i}$ & denotes the vertex that represents the physical initial station of trip~$i$ & - \\
			$\gamma_{i}$ & position of trip $i$ within a trip sequence & - \\
			$o_{k}^{\text{p}}$ & energy consumption rate for a bus of type $k$ with passengers on-board & kWh/km\\
			$o_{k}^{\text{e}}$ & energy consumption rate for a an empty bus of type $k$ & kWh/km\\
			$l_{i}$ & covered distance of operating trip $i$ & km\\
			$d_{ij}$ & distance between end of trip $i$ and start of subsequent trip $j$ & km\\
			$\tau_{ij}$ & time between end of trip $i$ and start of subsequent trip $j$ & h\\
			$t_{ij}$ & travel time between end of trip $i$ and start of subsequent trip $j$ & h\\
			\bottomrule
		\end{tabular}				
	\end{threeparttable}
	\centering
\end{table*}
We note that any set of bus types may contain several sub-types to account for different vehicle characteristics, e.g., heterogeneous battery capacities. Accordingly, we split the \glspl{abk:beb} into \gls{abk:beb-on} types~$\mathcal{K}^{\text{n}}$ and \gls{abk:beb-oc} types~$\mathcal{K}^{\text{o}}$ such that $\mathcal{K}^{\text{BEB}} = \mathcal{K}^{\text{n}} \cup \mathcal{K}^{\text{o}}$. 

Further, we denote with~$\mathcal{R}$ the set of potential locations where a charging facility can be built. These locations must be initial stations of bus lines, i.e., stations at which a trip starts. We use $\lambda_{i}\in\mathcal{R}$ to denote the vertex that represents the initial station of trip~$i$.

Let~$\mathcal{S}$ be the set of all trip sequences $s\in\mathcal{S}$ for a given bus schedule. We note that all trips are uniquely identified by a label $l\in\mathcal{L}$ such that trip sequences are formally disjoint. Accordingly, we represent a trip sequence~$s$ via two sets $\mathcal{N}_{s}$, $\mathcal{A}_{s}$. Here, $\mathcal{N}_{s}$ denotes the labels $l\in\mathcal{L}$ of all trips that belong to trip sequence~$s$, while set $\mathcal{A}_{s}$ denotes tuples~$(i,j)$, which denote that trip~$j$ succeeds trip~$i$ after either some dwell-time or dead-heading. We note that formally, $\mathcal{A}_{s}$ allows to construct~$\mathcal{N}_s$ and is thus sufficient to define~$s$, but the additional use of~$\mathcal{N}_s$ eases the notation of the \gls{abk:mip} model significantly. For a detailed explanation of the computation of the sets~$\mathcal{S}$, $\mathcal{N}_{s}$ and $\mathcal{A}_{s}$, which together define a vehicle schedule, we refer to Appendix~\ref{apx:A}.

We use decision variables~$n_{kt}$ to denote the number of buses of type~$k$ in period~$t$ in the bus fleet, and~$Q_{k}$ to denote the battery capacity of a bus of type $k\in\mathcal{K}^{\text{BEB}} $. Let~$h_{k}$ be the holding period of a bus of type~$k$, i.e., the time that a bus remains in the fleet before it it is sold. Decision variables~$p_{kt}$ define the number of buses of type~$k$ that are purchased at the beginning of period~$t$. Decision variables~$a_{t}$ depict the number of depot charging facilities in period~$t$. Binary decision variables~$y_{it}$ state whether a charging facility is available at bus station~$i$ in period~$t$. Binary decision variables~$x_{skt}$ indicate which bus type~$k$ is used to operate trip sequence~$s$ in period~$t$. We use variables~$\Theta_{st}$ as an artificial delimiter that denotes the minimum battery capacity which is necessary to operate trip sequence~$s$. For \glspl{abk:beb-oc}, $\Theta_{st}$ is limited by the net battery capacity~$\mu Q_{k}$. In contrast, for \glspl{abk:beb-on}, $\Theta_{st}$ is relaxed. With this delimiter, we formally decouple all energy balance constraints from being indexed to a specific bus type and keep the number of variables and constraints as small as  possible. Finally, decision variables~$q_{it}$ define the battery state of charge and~$w_{it}$ denotes the amount of energy recharged before operating trip~$i$ in period~$t$. With this notation our model holds as follows.
\paragraph{Objective}
Our Objective~(\ref{eq:obj}) minimizes the (discounted) \gls{abk:tco} associated with the integration of \glspl{abk:beb} into urban bus networks. It includes costs for the bus fleet transformation~($C^{\text{fleet}}$), the charging infrastructure installation~($C^{\text{infr}}$), and the bus network operation~($C^{\text{oper}}$).
Each of these cost types occurs on an annual basis: $C^{\text{fleet}}_{t}$, $C^{\text{infr}}_{t}$,  $C^{\text{oper}}_{t}$. Due to the finite planning horizon, the fleet operator needs to purchase the initial bus fleet for its initial value~$C^{\text{fleet}}_{0}$. To avoid unintended effects of the finite planning horizon, we account for the salvage value~$V^{\text{fleet}}_{\text{n+1}}$ of the bus fleet at the end of the planning horizon.

{\footnotesize
\vspace{-0.25cm}
\begin{multline}
\text{min} ~ Z =  TCO = C^{\text{fleet}} + C^{\text{infr}} + C^{\text{oper}} =
C^{\text{fleet}}_{0} 
+
\sum_{t \in \mathcal{T}}{
\frac{C^{\text{fleet}}_{t}+C^{\text{infr}}_{t}+C^{\text{oper}}_{t}}{(1+p)^{t}}
}
- \frac{V^{\text{fleet}}_{\text{n}+1}}{(1+p)^{\text{n}+1}}\hfill
\label{eq:obj}
\end{multline}}
The (annual) costs associated with the bus fleet transformation~($C^{\text{fleet}}_{t}$) described in Equation~(\ref{eq:fleet}) consist of costs for purchasing new buses~($C^{\text{bus.purch}}_{t}$), purchasing batteries and replacing them after their economic lifetime $e^{q}$~($C^{\text{bat.purch}}_{t}$), minus revenues~($V^{\text{bus.sold}}_{t}$) due to selling buses for a salvage value at the end of the holding period $h_{k}$. The salvage value at the end of the planning period is determined by a linear depreciation of the purchase costs.
	
{\footnotesize
	\vspace{-0.25cm}
	\setlength{\abovedisplayskip}{3pt}
	\setlength{\abovedisplayshortskip}{3pt}
	\setlength{\belowdisplayskip}{3pt}
	\setlength{\belowdisplayshortskip}{3pt}
	\begin{multline}
	\begin{aligned}
	C^{\text{fleet}}_{t} &= C^{\text{bus.purch}}_{t} + C^{\text{bat.purch}}_{t} - V^{\text{bus.sold}}_{t}\\
	&=\summe{k \in \mathcal{K}} c_{k}^{\text{b}} p_{kt}
	+
	\summe{k \in \mathcal{K}^{\text{BEB}}} c_{kt}^{\text{q}} Q_{k} \left( p_{kt}+p_{k,t-e^{\text{q}}} \right)
	-
	\summe{k \in \mathcal{K}} \left( \left( c_{k}^{\text{b}} - v^{\text{b}} \right)\left( 1 - \frac{h_{k}}{e^{\text{b}}} \right) + v^{\text{b}} \right) p_{k,t-h_{k}}
	\end{aligned}\hfill \forall t \in \mathcal{T}
	\label{eq:fleet}
	\end{multline}}
	Costs for charging infrastructure~($C^{\text{infr}}$) consist of installation and maintenance costs, as described in Equation~(\ref{eq:C_f}). Costs for installing charging infrastructure~($C^{\text{infr.install}}_{t}$) only occur for newly installed charging facilities, i.e., if binary variable $y_{it}$ changes its value from zero to one, or if integer variable $a_{t}$ increases, while maintenance costs~($C^{\text{infr.maint}}_{t}$) occur for all charging facilities in all time steps.
	
	{\footnotesize
	\vspace{-0.25cm}
	\setlength{\abovedisplayskip}{3pt}
	\setlength{\abovedisplayshortskip}{3pt}
	\setlength{\belowdisplayskip}{3pt}
	\setlength{\belowdisplayshortskip}{3pt}
	\begin{multline}
	\begin{aligned}
	C^{\text{infr}}_{t} &= C^{\text{infr.install}}_{t} + C^{\text{infr.maint}}_{t} \\
	&=c^{\text{o}} \sum_{i \in \mathcal{R}}
	\left( y_{it} - y_{i,t-1} \right)
	+ c^{\text{n}}(a_{t}-a_{t-1}) 
	+ c^{\text{om}} \sum_{i \in \mathcal{R}} y_{it} +
	c^{\text{nm}}a_{t}
	\end{aligned}
	\hfill \forall t \in \mathcal{T}
	\label{eq:C_f}
	\end{multline}}
	As described in Equation~(\ref{eq:oper}), operational costs~($C^{\text{oper}}$) occur due to maintenance and repair of the bus fleet~($C^{\text{oper.maint}}_{t}$) as well as due to energy consumption~($C^{\text{oper.energy}}_{t}$), both depending on the selected bus type~($x_{skt}$) and the total distance covered. Herein, we consider heterogeneous energy consumptions for service trips with passengers~($l_{i}$) and deadheading trips without passengers~($d_{ij}$).
	
	{\footnotesize
	\vspace{-0.25cm}
	\setlength{\abovedisplayskip}{3pt}
	\setlength{\abovedisplayshortskip}{3pt}
	\setlength{\belowdisplayskip}{3pt}
	\setlength{\belowdisplayshortskip}{3pt}
	\begin{multline}
	\begin{aligned}
	C^{\text{oper}}_{t} &= \eta \left( C^{\text{oper.maint}}_{t} + C^{\text{oper.energy}}_{t}  \right)
	\\
	&=\eta \left( 
	\sum_{k \in \mathcal{K}}\sum_{s \in \mathcal{S}} c_{k}^{\text{bm}}
	\left(
	\sum_{i \in \mathcal{N}_{s}}l_{i} +
	\summe{(i,j) \in \mathcal{A}_{s}}d_{ij}
	\right)
	x_{skt}
	+
	\sum_{k \in \mathcal{K}}\sum_{s \in \mathcal{S}}
	c_{k}^{\text{e}} \left(
	o_{k}^{\text{p}} \sum_{i \in \mathcal{N}_{s}}l_{i} +
	o_{k}^{\text{e}} \summe{(i,j) \in \mathcal{A}_{s}}d_{ij}
	\right)
	x_{skt}	
	\right)
	\end{aligned}
	\hfill \forall t \in \mathcal{T}
	\label{eq:oper}
	\end{multline}}
	At the end of the planning horizon, we take a payment~($V^{\text{fleet}}_{\text{n}+1}$) equal to the total final salvage value of all buses~($V^{\text{bus.salv}}_{\text{n}+1}$) and batteries~($V^{\text{bat.salv}}_{\text{n}+1}$) into account. Again, the salvage value results from a linear depreciation, as described in Equation~(\ref{eq:final_fleet}).

{\footnotesize
\vspace{-0.25cm}
\setlength{\abovedisplayskip}{3pt}
\setlength{\abovedisplayshortskip}{3pt}
\setlength{\belowdisplayskip}{3pt}
\setlength{\belowdisplayshortskip}{3pt}
\begin{multline}
\begin{aligned}
V^{\text{fleet}}_{\text{n}+1} &= V^{\text{bus.salv}}_{\text{n}+1} + V^{\text{bat.salv}}_{\text{n}+1} \\
&=\summe{k \in \mathcal{K},\ t \in \lbrace \mathcal{T}\,|\,t \geq t_{\text{n}+1} - h_{k} \rbrace}
\left( \left( c_{k}^{\text{b}} - v^{\text{b}}\right) \left( 1 - \frac{t_{\text{n+1}}-t}{e^{\text{b}}} \right) + v^{\text{b}} \right) p_{kt}
+
\summe{k \in \mathcal{K}^{\text{BEB}},\ t \in \lbrace \mathcal{T}\,|\,t \geq t_{\text{n+1}} - e^{\text{q}} \rbrace}
c_{kt}^{\text{q}} Q_{k} \left( 1 - \frac{t_{\text{n+1}}-t}{e^{\text{q}}} \right) (p_{kt}+p_{k,t-e^{\text{q}}})
\end{aligned}
\hfill
\label{eq:final_fleet}
\end{multline}}
\paragraph{Constraints}
Our objective is subject to the following constraints.
{\footnotesize
\vspace{-0.25cm}
\setlength{\abovedisplayskip}{3pt}
\setlength{\abovedisplayshortskip}{3pt}
\setlength{\belowdisplayskip}{3pt}
\setlength{\belowdisplayshortskip}{3pt}
\begin{multline}
n_{kt} = n_{k,t-1} + p_{kt} - p_{k,t-h_{k}}
\hfill \forall k \in \mathcal{K}, t \in \mathcal{T}
\label{eq:stock}
\end{multline}
\begin{multline}
a_{t} \geq a_{t-1}
\hfill \forall t \in \mathcal{T}
\label{eq:aa}
\end{multline}
\begin{multline}
y_{it} \geq y_{i,t-1}
\hfill \forall i \in \mathcal{R}, t \in \mathcal{T}
\label{eq:yy}
\end{multline}
\begin{multline}
\sum_{k \in \mathcal{K}}x_{skt} = 1
\hfill \forall s \in \mathcal{S}, t \in \mathcal{T}
\label{eq:x}
\end{multline}
\begin{multline}
n_{kt} \geq \sum_{s \in \mathcal{S}} x_{skt}
\hfill \forall k \in \mathcal{K}, t \in \mathcal{T}
\label{eq:f}
\end{multline}
\begin{multline}
\left( o_{k}^{\text{p}} \sum_{i \in \mathcal{N}_{s}}l_{i} +
o_{k}^{\text{e}} \summe{(i,j) \in \mathcal{A}_{s}}d_{ij}
\right) x_{skt} \leq \mu Q_{k}
\hfill \forall s \in \mathcal{S}, k \in \mathcal{K}^{\text{n}}, t \in \mathcal{T}
\label{eq:ON}
\end{multline}
\begin{multline}
\Theta_{st} \leq \mu Q_{k} + M (1-x_{skt})
\hfill \forall s \in \mathcal{S}, k \in \mathcal{K}^{\text{o}}, t \in \mathcal{T}
\label{eq:Qst}
\end{multline}
\begin{multline}
0 \leq q_{it} \leq \Theta_{st} - d_{0i} o_{\text{BEB}}^{\text{e}}
\hfill \forall s \in \mathcal{S}, i \in \lbrace \mathcal{N}_{s} \vert \gamma_{i}=0 \rbrace, t \in \mathcal{T}
\label{eq:0}
\end{multline}
\begin{multline}
0 \leq q_{jt} \leq q_{it} + w_{it} - l_{i}o_{\text{BEB}}^{\text{p}} - d_{ij}o_{\text{BEB}}^{\text{e}}
\hfill \forall s \in \mathcal{S}, (i,j) \in \lbrace \mathcal{A}_{s} \vert  \gamma_{i} \neq 0 \rbrace , t \in \mathcal{T}
\label{eq:A}
\end{multline}
\begin{multline}
0 \leq w_{jt} \leq  r (\tau_{ij} - t_{ij}) y_{\lambda_{j}t}
\hfill \forall s \in \mathcal{S}, (i,j) \in \mathcal{A}_{s}, t \in \mathcal{T}
\label{eq:w}
\end{multline}
\begin{multline}
w_{it} \leq \Theta_{st} - q_{it}
\hfill \forall s \in \mathcal{S}, i \in \mathcal{N}_{s}, t \in \mathcal{T}
\label{eq:N}
\end{multline}
\begin{multline}
n_{kt}, p_{kt} \in \mathbb{N}
\hfill \forall k \in \mathcal{K}, t \in \mathcal{T}
\label{eq:def:n}
\end{multline}
\begin{multline}
a_{t} \in \mathbb{N}
\hfill \forall t \in \mathcal{T}
\label{eq:def:a}
\end{multline}
\begin{multline}
y_{it} \in \lbrace 0;1 \rbrace
\hfill \forall i \in \mathcal{R}, t \in \mathcal{T}
\label{eq:def:y}
\end{multline}
\begin{multline}
x_{skt} \in \lbrace 0;1 \rbrace
\hfill \forall s \in \mathcal{S}, k \in \mathcal{K}, t \in \mathcal{T}
\label{eq:def:x}
\end{multline}}
Constraints~(\ref{eq:stock}) define the holding balance for each bus type and thus the point in time when buses of a certain type must be sold. Constraints~(\ref{eq:aa}) and~(\ref{eq:yy}) ensure that an installed charging facility cannot be uninstalled in a subsequent period. Constraints~(\ref{eq:x}) ensure that any trip sequence~$s$ is served by exactly one bus type~$k$. To guarantee operational feasibility, Constraints~(\ref{eq:f}) enforce that for each bus type $k$ the number of buses available is sufficient for the fleet's operation at any point in time. In order to ensure the bus network operation with limited driving ranges of \glspl{abk:beb}, Constraints~\mbox{(\ref{eq:ON})\,--\,(\ref{eq:N})} preserve energy balance constraints and secure that a bus does not run out of energy during operations. Here, Constraints~(\ref{eq:ON}) ensure that the total energy consumption of a trip sequence does not exceed the net available battery capacity~$\mu Q_{k}$ if an \gls{abk:beb-on} is assigned to it. Constraints~\mbox{(\ref{eq:Qst})\,--\,(\ref{eq:N})} ensure that an assigned \gls{abk:beb-oc} never runs out of energy, while accounting for (partial) recharging operations. Specifically, Constraints~(\ref{eq:Qst}) limit the artificial delimiter $\Theta_{st}$ to the net available battery capacity~$\mu Q_{k}$. By linking the artificial delimiter $\Theta_{st}$ with the binary sequence assignment variable $x_{skt}$ in Constraints~(\ref{eq:Qst}), we formally decouple the assignment decisions from the following energy balance Constraints~(\ref{eq:0})\,--\,(\ref{eq:N}) to reduce the model complexity. Constraints~(\ref{eq:0}) limit the battery state of charge~$q_{it}$ before operating the first trip of a trip sequence, and Constraints~(\ref{eq:A}) indicate the battery state of charge~$q_{jt}$ ahead of all remaining trips and at depot arrival. Since the amount of recharged energy depends on recharging time and charging power, Constraints~(\ref{eq:w}) limit the recharged energy~$w_{it}$ to the product of net available time $\tau_{ij}-t_{ij}$ and the installed charging power~$r$. Constraints~(\ref{eq:N}) ensure that the recharged energy~$w_{it}$ does not exceed the remaining battery capacity $Q_{k}-q_{it}$. We note that Constraints~(\ref{eq:Qst})\,--\,(\ref{eq:N}) only affect \glspl{abk:beb-oc}, since~$\Theta_{st}$ has a dummy status if~$x_{skt}=1, k \notin \mathcal{K}^{\text{o}}$. Therefore, for \glspl{abk:beb-on}, Constraints~(\ref{eq:Qst})\,--\,(\ref{eq:N}) do not apply, as $\Theta_{st}$ remains a sufficiently high value so that the related energy balance constraints are relaxed.  Finally, Constraints~(\ref{eq:def:n})\,--\,(\ref{eq:def:x}) state integer and binary domains.
\section{Case Study}\label{sect:caseStudy}
\begin{figure}[!hb]
	\centering
	\begin{minipage}{\textwidth}
		\centering
		\begin{tikzpicture}[scale=1.175]
		\node[inner sep=0pt] at (3.4,2.85)
		{\includegraphics[scale=0.425]{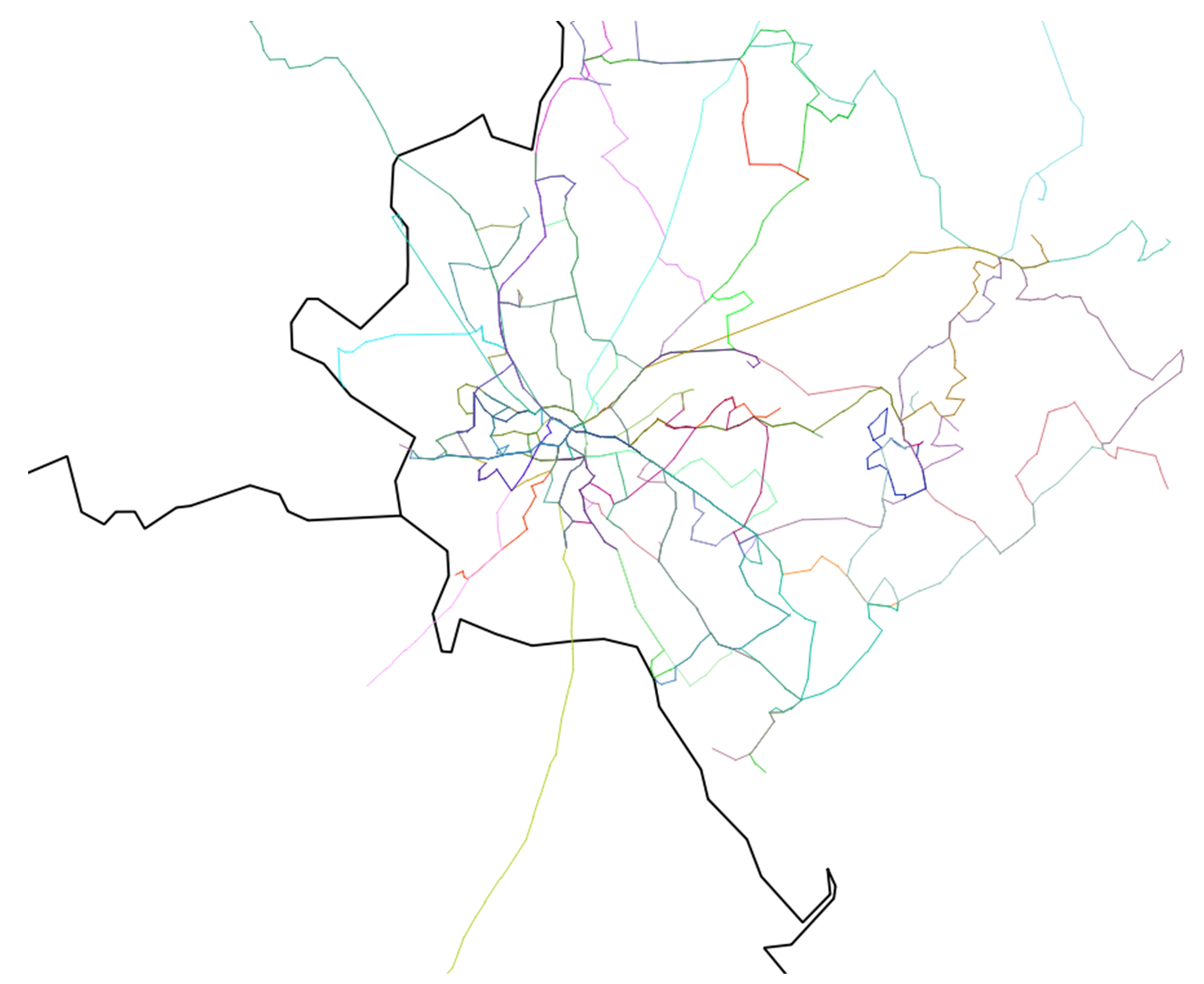}};
		\begin{axis}[%
	xmin=0,
	xmax=100,
	xticklabels={,,},
	xtick style={draw=none},
	ymin=0,
	ymax=100,
	yticklabels={,,},
	ytick style={draw=none},
	legend style={at={(0,0)},anchor=south west,font=\footnotesize},
	legend cell align={left},
	]
	\node at (15,75) {\footnotesize Netherlands};
	\node at (15,25) {\footnotesize Belgium};
	\node at (85,25) {\footnotesize Germany};
	\addplot[mark=*, only marks] coordinates {(46,56)};
	\addlegendentry{center of Aachen};

\end{axis}
		\end{tikzpicture}
	\end{minipage}
	\caption{Bus network in Aachen.}
	\label{fig:aseag}
\end{figure}
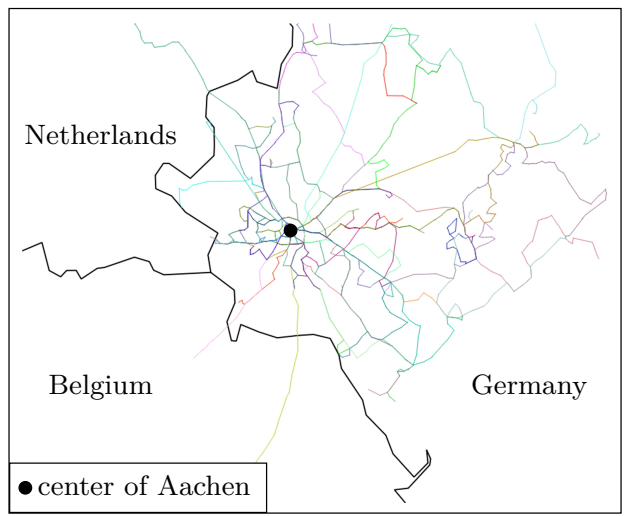

We base our case study on the real-world bus network in the city of Aachen, Germany~(see Figure~\ref{fig:aseag}), which is one of the biggest European public transport systems that relies exclusively on bus transportation. Within one year, the buses of the network's operator \gls{abk:aseag} cover a distance of~$20$ million kilometers and transport more than~$71$ million passengers resulting in a transport volume of~$101$ million passenger kilometers per year~(\citealt{ASEAG2017}). In total,~$55,9$ million Euros of bus fares were captured in 2019. 
To serve~$108$ bus lines, $498$ buses are available and~$600$ bus drivers are employed. Besides regular bus lines, this includes school buses, night buses, express buses and dial-a-bus services. In total, the buses travel on average~$56,000$\,km per day. The average distance of bus lines respective trips is between~$3.2$ and~$41$\,km, while the length of a sequence a bus is driving per day~(i.e., the sum of all trips and dead-heading activities) lies between~$40$\,km and~$481$\,km. \gls{abk:aseag} operates this fleet with a single depot.

Currently, legal \gls{abk:nox} concentration thresholds are exceeded in the city center of Aachen. Accordingly, local public transport authorities aim at operating public transport and thus the \gls{abk:aseag}'s bus fleet mainly by \glspl{abk:beb} in the future~(\citealt{LRP2018}). As a result, the local bus operator \gls{abk:aseag} currently considers to comprehensively integrate \glspl{abk:beb} into its fleet, herein facing the need to provide sufficient charging facilities and to ensure operational feasibility. Given the high heterogeneity of bus lines regarding their different lengths and number of bus stops, the \gls{abk:aseag} bus network constitutes a promising case study to observe a multitude of structural effects when applying an integrated planning approach.

Accordingly, we base our analysis on the transformation process of the \gls{abk:aseag} bus network towards \glspl{abk:beb}. We analyze the underlying network with the regular bus lines, i.e., we omit the school, night, express, and dial-a-service lines. In total, we consider~$60$ regular bus lines with~$5,092$ trips per workday serving~$1,018$ bus stops~(\citealt{AVV2016}). In the following, we detail the used data and our methodology to evaluate \gls{abk:nox} emissions in Section~\ref{sub:caseStudy-data}, before we describe our experimental design in Section~\ref{sub:caseStudy-design}. 
\subsection{Data}\label{sub:caseStudy-data}
We consider a time horizon of~$20$ years, $\mathcal{T}=\{2020,\,2021,\,...,\,2039\}$ and a fleet size that equals the number of necessary trip sequences, such that one bus is available for each trip sequence. To cover the considered~$60$ bus lines, the fleet requires $|\mathcal{S}|=357$ buses. 
 
We assume a maximum holding period of buses, i.e., a vehicle lifetime, of $12$ years. Accordingly, we introduce $|\mathcal{K}^{\text{ICEB}}|=12$ different \gls{abk:iceb} classes, each with a specific holding period $h_{k}\in\{1,...,12\}$, to allow for a heterogeneous age mix of \glspl{abk:iceb} in the fleet. For \glspl{abk:beb} we assume that once purchased all vehicles are kept for the maximum holding period of 12 years.

We consider different battery capacities $Q_k \in\{100,200,300,400\}$\,kWh that reflect the range of technical specifications currently offered on the market~(cf. \citealt{GaoEtAl2017}), with $Q_{\text{max}}=400$\,kWh being the maximum available battery capacity. Also, we consider \glspl{abk:beb-on} that are only charged overnight at the depot, and \glspl{abk:beb-oc} that can be additionally charged during operations. Combining the different battery capacity classes and charging concepts, we account for $|\mathcal{K}^{\text{n}}|=4$ types of \glspl{abk:beb-on} and $|\mathcal{K}^{\text{o}}|=4$ types of \glspl{abk:beb-oc}.  

We allow any initial bus station of a trip to serve as a potential opportunity charging facility location such that the set of potential opportunity charging facilities consists of~$|\mathcal{R}|=182$ locations.  Table~\ref{tab:technical_data} details further technical data that is necessary to specify all parameters. If a bus performs a dead-heading operation without passengers on board, we reduce its consumption without passengers on-board to $75$\,\% of the consumption with passengers on-board.

We base our cost term estimates data from recent field projects and expert knowledge. Table~\ref{tab:cost_terms} provides an overview of these cost terms and their references. 
\begin{table}[!hb]
	\centering
	\begin{threeparttable}
		\centering
		\setlength{\abovecaptionskip}{0.5ex}
		\addtolength{\tabcolsep}{-3pt}	
		\caption{Technical data.}
		\label{tab:technical_data}
		\footnotesize
		\begin{tabular}{rllll}
			\toprule
			parameter								& description												& value				& unit	& reference\\
			\midrule
			$e$										& economic lifetime of any bus								& 12				& years	& II, III, IV \\
			$h_{k}$									& holding period of a bus of type $k$						& 1,\,...,\,12		& years	&  \\
			$Q_{k}$									& battery capacities							& 100,\,...,\,400	& kWh	& V \\
			$\mu$									& usable percentage of battery capacity						& 80				& \%	& II, IV \\
			$b$										& economic lifetime of a battery							& 6					& years	& II\\	
			$o_{\text{\gls{abk:iceb}}}^{\text{f}}$	& diesel consumed by an \gls{abk:iceb} 	& 0.61				& l/km	& I\\
			$o_{\text{\gls{abk:beb}}}^{\text{f}}$	& electricity consumed by a \gls{abk:beb} 	& 2.06				& kWh/km & III\\
			\bottomrule
		\end{tabular}
		\begin{tablenotes}
			\scriptsize\item	
			(I)\,--\,(V) refer to references as follows:
			(I)\,\citealt{ASEAG2017}, (II)\,\citealt{IKT2015}, (III)\,\citealt{Lajunen2018}, (IV) \citealt{Rogge2018}, (V) \cite{GaoEtAl2017}.
		\end{tablenotes}
	\end{threeparttable}
\end{table}
We note that cost terms of opportunity charging facilities and batteries for \glspl{abk:beb-oc} depend on charging power. To determine cost values for battery capacities and charging facility technologies that are not listed in Table~\ref{tab:cost_terms}, we apply a linear interpolation between listed values. For night charging facilities and \glspl{abk:beb-on}, we assume a charging power of $50$\,kW. We anticipate operational costs by calculating a number of workday equivalent days, i.e., we convert weekend and public holiday timetables into their workday equivalent.

Table~\ref{tab:emissions} summarizes the initial bus fleet composition~(see \citealt{LRP2018}) and the legal \gls{abk:nox} emission thresholds~(see \citealt{EU2009}) that these buses comply with. Regarding the fleet composition at the start of the planning horizon in $t_0$, we assume that all buses comply with the \gls{abk:nox} threshold that was in force when the bus was purchased, i.e., emissions depend on the age of the bus.

When an \gls{abk:iceb} is purchased  during the planning horizon, we assume that it complies with the newest EU-VI standard. For \glspl{abk:beb}, we assume zero local \gls{abk:nox} emissions. Further, we assume that the operator replaces older \glspl{abk:iceb} with high emissions first. We then calculate \gls{abk:nox} emissions a-posteriori converting EURO emission thresholds for heavy-duty vehicles and buses~$E_{\text{NO}_{\text{x}}}$ that are defined in~kWh to mass emissions per kilometer~$\widehat{E_{\text{NO}_{\text{x}}}}$ as
$$
\widehat{E_{\text{NO}_{\text{x}}}} \scriptsize{\bigg[\frac{\text{g}}{\text{km}}\bigg]}
= 
E_{\text{NO}_{\text{x}}} \scriptsize{\bigg[\frac{\text{g}}{\text{kWh}}\bigg]}
o \scriptsize{\bigg[\frac{\text{m}^{3}}{\text{km}}\bigg]}
\rho \scriptsize{\bigg[\frac{\text{kg}}{\text{m}^{3}}\bigg]}
H \scriptsize{\bigg[\frac{\text{kWh}}{\text{kg}}\bigg]}	,
\hfill
$$
with  $o = 0.00059$, $\rho = 832.5$, and $H = 11.9$.
\subsection{Design of Experiments}\label{sub:caseStudy-design}
We split our experiments into two parts. We first perform an a-priori feasibility study to discuss structural characteristics  of the case study. We then perform optimization based analyses by solving the \gls{abk:mip} as introduced in Section~\ref{sect:model}.
\begin{table}[!hb]
	\centering
	\begin{threeparttable}
		\centering
		\setlength{\abovecaptionskip}{0.5ex}
		\addtolength{\tabcolsep}{-3pt}	
		\caption{Cost terms.}
		\label{tab:cost_terms}
		\footnotesize
		\begin{tabular}{rllll}
			\toprule
			parameter								& description									& value		& unit			& reference\\
			\midrule
			$c_{{\text{\gls{abk:iceb}}}}^{\text{b}}$& purchase costs of an \gls{abk:iceb}			& 330,000	& \euro			& I	\\
			$c_{{\text{\gls{abk:beb}}}}^{\text{b}}$	& purchase costs of \gls{abk:beb} (w/o battery)	& 350,000	& \euro			& I*	\\			
			$v^{\text{b}}$							& salvage value of purchase costs of a bus		& 7			& \% of $c_{\text{\gls{abk:iceb}}}^{\text{b}}$ & I\\
			$c_{k0}^{\text{q}}$						& initial battery costs (50\,kW)				& 487.5		& \euro/kWh		& III**\\
			$c_{k0}^{\text{q}}$						& initial battery costs (350\,kW)				& 780		& \euro/kWh		& III**\\
			$c_{\text{\gls{abk:iceb}}}^{\text{bm}}$	& maintenance costs of an \gls{abk:iceb}		& 0.5		& \euro/km		& I	\\
			$c_{\text{\gls{abk:beb}}}^{\text{bm}}$	& maintenance costs of a \gls{abk:beb}			& 0.44		& \euro/km		& I	\\
			$c_{\text{\gls{abk:iceb}}}^{\text{e}}$	& diesel price									& 0.97		& \euro/l		& IV\\
			$c_{\text{\gls{abk:beb}}}^{\text{e}}$	& electricity price								& 0.13		& \euro/kWh 	& IV\\	
			$c^{\text{o}}$							& costs of an opportunity charging facility (50\,kW) & 30,000	& \euro			& V	\\
			$c^{\text{o}}$							& costs of an opportunity charging facility (150\,kW) & 90,000	& \euro			& VI\\
			$c^{\text{o}}$							& costs of an opportunity charging facility (350\,kW)& 134,250	& \euro			& VI\\
			$c^{\text{n}}$							& costs of a night charging facility (depot) 	& 5,000		& \euro			& V	\\
			$c^{\text{om/nm}}$						& maintenance costs of a charging facility		& 1			& \%/year of $c^{\text{o/n}}$ & V\\	
			$g$										& interest rate									& 5			& \%/year			& I	\\
			$\eta$									& annual days of operation (fictional)			& 307		& days/year		& \\
			\bottomrule
		\end{tabular}
		\begin{tablenotes}
			\scriptsize		
			\item
			(I)\,--\,(VI) characterize references as follows:
			(I)\,\citealt{IKT2015}, (II)\,\citealt{KunithBook}, (III)\,\citealt{Lajunen2018}, (IV)\,\citealt{Destatis2018}, (V)\,\citealt{NPE2015}, (VI)\,expert knowledge.
			\item[*] adapted based on specific bus type characteristics.
			\item[**] updated considering annual battery price reduction ($2.5$\,\%).
		\end{tablenotes}
	\end{threeparttable}
\vspace{-0.3cm}
\end{table}
\begin{table}[!hb]
	\centering
	\begin{threeparttable}
		\centering
		\setlength{\abovecaptionskip}{0.5ex}
		\addtolength{\tabcolsep}{-3pt}	
		\caption{Emission factors.}
		\label{tab:emissions}
		\footnotesize
		\begin{tabular}{llll} 	
			\toprule
			number of buses\tnote{*}	& standard\tnote{**} 	& $E_{\text{NO}_{\text{x}}}$\,[g/kWh] 	& $\widehat{E_{\text{NO}_{\text{x}}}}$\,[g/km]\\
			\midrule
			109		& EU-III					& 5.0									& 29.22	\\		
			165		& EU-V\,/\,EEV				& 2.0									& 11.69	\\
			83		& EU-VI						& 0.4									& 2.38	\\
			\bottomrule
		\end{tabular}
		\begin{tablenotes}				
			\scriptsize
			\item[*] \citealt{LRP2018},\textsuperscript{**}\citealt{EU2009}.
		\end{tablenotes}	
	\end{threeparttable}
\vspace{-0.3cm}
\end{table}

\noindent \textit{A-priori feasibility study:}
We perform an a-priori feasibility study to analyze the minimum battery capacities that are necessary for an electrification of the bus network dependent on the available charging power. Additionally, we analyze the maximum share of \glspl{abk:beb} that can be feasibly deployed for a specific battery capacity and charging power. These results allow for in depth analyses of the optimization-based analyses with regard to its solution space, i.e., allow to quantify whether a certain share of \glspl{abk:beb} results from economic conditions or technological limitations.

\noindent \textit{Optimization-based analyses:}
We define each scenario by its purchasable bus types. Table~\ref{tab:design} summarizes all possible scenarios, including a reference scenario in which all different bus types are available~(\texttt{all}), a scenario in which only \glspl{abk:iceb} are available~(\texttt{ic}), as well as two scenarios in which either only \glspl{abk:iceb} and \glspl{abk:beb-on}~(\texttt{nc}) or only \glspl{abk:iceb} and \glspl{abk:beb-oc}~(\texttt{oc}) are available.

For each scenario, we analyze different parameter settings for the available charging power~($r$) and annual battery price reductions~($b$). We vary the available charging power~$r$ in \mbox{$r\in[50,\,100,\,...,\,500]$\,kW} and consider battery price reductions \mbox{$b\in[0,\,2.5,\,...,\,12.5]$\,$\%$/year}. With these parameter variations, we capture an ongoing discussion between academics and practitioners whether high capacity batteries should be favored over high power charging facilities combined with small-capacity batteries or vice versa. Concluding, we consider four scenarios~(\texttt{all}, \texttt{ic}, \texttt{nc}, \texttt{oc}) and~$60$ different parameter settings (\texttt{r\footnotesize{0-500}\normalsize{|b}\footnotesize{0-12.5}}). 

We identify each scenario by a unique identifier, stating~(in this order) the bus type scenario, the available charging power, and the considered battery price reduction, e.g., our basic scenario \texttt{(all|r\footnotesize{350}\normalsize{|b}\footnotesize{2.5}}) refers to a scenario where all bus types are available, the available charging power of charging facilities is $r=350$\,kW, and the annual battery price reduction is $b=2.5$\,\%/year. We split our analyses in two parts, first analyzing a base case, motivated from our a-priori feasibility studies, and a subsequent sensitivity analyses over the stated parameter ranges.
\section{Results}\label{sect:results}
In the following, we first discuss our a-priori feasibility study in Section~\ref{sub:feasibility}. Afterwards, we discuss results for the reference scenario in Section~\ref{sub:baseCase}, before we detail the scenario analysis in Section~\ref{sub:scenarioAnalysis}, and derive managerial insights in Section~\ref{sub:insights}. We solved the \gls{abk:mip} to optimality for all cases within a computational time limit of 60 minutes, using a standard desktop computer with 16-GB-RAM and 3.6 ghz (Intel i7-4790) using Gurobi-8.1.0.
\subsection{A-priori Feasibility Study}\label{sub:feasibility}
With this a-priori analysis we aim at identifying the space of operationally feasible solutions, i.e., we identify which sequences can be operated by a \gls{abk:beb} dependent on the available power at charging facilities and dependent on the battery capacity of the \glspl{abk:beb}. To do so, we determine for each sequence \mbox{$s\in\mathcal{S}$} the minimum battery capacity that is necessary to operate a \gls{abk:beb} on this sequence.
For these analyses, we consider the most optimistic scenario and assume that an \gls{abk:frf} is available at any first or final station of a trip. To check the feasibility of a sequence, we then use Constraints~(\ref{eq:0})\,--\,(\ref{eq:w}). We repeat this analysis for different power levels at charging facilities to explore the interdependency between the available charging power level and the available battery capacity.
\begin{table*}[!hb]
	\centering
	\begin{threeparttable}
		\centering
		\setlength{\abovecaptionskip}{0.5ex}
		\addtolength{\tabcolsep}{-3pt}
		\caption{Bus types per scenario.}
		\label{tab:design}
		\footnotesize
		\def\c{\checkmark}
		\begin{tabular}{rccc}		
			\toprule
			& \multicolumn{3}{c}{bus type}\\
			\cmidrule(lr{1em}){2-4}	
			scenario& \gls{abk:iceb} & \gls{abk:beb-on} & \gls{abk:beb-oc} \\
			\midrule
			\texttt{all}& \c & \c & \c \\
			\texttt{ic} & \c &    &    \\
			\texttt{nc} & \c & \c &    \\
			\texttt{oc} & \c &    & \c \\
			\bottomrule		
		\end{tabular}
		\centering
	\end{threeparttable}
\vspace{-0.7cm}
\end{table*}

Figure~\ref{fig:feasibility-1} shows a Box-Whisker-Plot that details the distribution of necessary battery capacities to operate all~$|\mathcal{S}|=357$ sequences with \glspl{abk:beb} at a certain charging power level. Further, the figure indicates the maximum battery capacity~$Q_{\text{max}}=400$\,kWh with a dashed line. As can be seen, a complete fleet transformation that allows to operate all sequences with \glspl{abk:beb} is only feasible with either battery capacities or charging power levels that exceed the current technological state of the art. 

Figure~\ref{fig:feasibility-2} shows the share of sequences~$s \in \mathcal{S}$, which can be operated with a \gls{abk:beb} at a certain charging power level, if we set the available battery capacity to its maximum value.
As can be seen, an electrification of~$62.5$\,\% to~$93.3$\,\% of the total fleet can be realized with current status-quo charging power levels between~$50$\,kW and~$150$\,kW.
For higher shares of electrification, one has to significantly increase the charging power level. High-power chargers of the latest generation with a charging power of~$350$\,kW allow for the electrification of~$99.72$\,\% of all trips. To achieve a full electrification, a charging power of at minimum~$400$\,kW would be necessary. Concluding, we note that a maximum electrification of~$99.72$\,\% can be reached in our reference scenario~(\texttt{all|r\footnotesize{350}\normalsize{|b}\footnotesize{2.5}}).
\subsection{Reference Scenario}\label{sub:baseCase}
Figure~\ref{fig:fleet} shows the cost-optimal bus fleet composition over the planning horizon $\mathcal{T}= \lbrace 2020,\,...,\,2039 \rbrace$ and its corresponding \gls{abk:tco} balance that results from solving the \gls{abk:mip} from Section~\ref{sect:model} for reference scenario~(\texttt{all|r\footnotesize{350}\normalsize{|b}\footnotesize{2.5}}). Here, the subscript of each \gls{abk:beb} type indicates its battery capacity, e.g., $\text{\gls{abk:beb-oc}}_{100}$ denotes \glspl{abk:beb-oc} with a battery capacity of~$100$\,kWh.
As can be seen, the fleet transformation evolves over time with a steadily increasing share of \glspl{abk:beb-oc}. Remarkably, only \glspl{abk:beb-oc} are integrated into the fleet, mostly with battery capacities of $100$\,kWh, $200$\,kWh or $300$\,kWh. Only two \glspl{abk:beb-oc} with the maximum capacity of~$400$\,kWh are integrated in the year~$2031$. Overall, the realized deployment of \glspl{abk:beb} falls short compared to its feasible maximum of~$99.72$\% as only~$83$\,\% of all buses are replaced by \glspl{abk:beb} throughout the planning horizon. Obviously, the remaining buses are not replaced for economic reasons. This steady state is already reached after the first half of the planning horizon in~$2031$ such that the operational cost savings of the \gls{abk:beb} can compensate the higher investment costs.

Our results show that the cost-optimal transformation takes place stepwise, which may appear counter-intuitive on the first glimpse. Analyzing this effect shows that each bus sequence has an individual break-even point that may differ over time due to additional effects, e.g., 
\begin{figure}[!hb]
	\centering
	\hspace{-0.5cm}
	\begin{minipage}[t]{0.45\textwidth}
		\centering
		\input{content/results/scenarioAnalysis/feasibility/feasibility.tex}
		\vspace{-0.25cm}
		\caption{Box-Whisker-Plot indicating minimum required battery capacity of any sequence for a variation of charging power \texttt{\footnotesize{(r50-500|bX)}}.}
		\label{fig:feasibility-1}
	\end{minipage}
	\hspace{0.5cm}
	\begin{minipage}[t]{0.45\textwidth}
		\centering
		\pgfplotstableread[col sep = semicolon]{content/results/scenarioAnalysis/feasibility/feasibility_Q_max.csv}\DataForBars 
		\begin{tikzpicture}[scale=1]
\begin{axis}[
	label style={font=\small},
	tick label style={font=\small},
	tick pos=left,
	y label style={at={(axis description cs:.03,.5)},anchor=south},
	ylabel near ticks,
	xmin=1,xmax=549,
	legend cell align={left},
	grid=major,
	ytick pos=left,
	ylabel={feasible share of \glspl{abk:beb} [\%]},
	ylabel near ticks,
	xtick distance=50,
	tick style = {black},
	axis on top=true,
	xtick pos=left,
	xlabel={charging power level [kW]},
	label style={font=\small},
	tick label style={font=\small},
]
	\addplot[mark=x,black,only marks,mark size=3.5,mark options = {thick}] table [x=r,y=share]  {\DataForBars};	
\end{axis}
\end{tikzpicture}
		\vspace{-0.25cm}
		\caption{Feasible share of \glspl{abk:beb} for a variation of charging power \texttt{\footnotesize{(r50-500|bX)}}, assuming a maximum battery capacity of $400$\,kWh.}
		\label{fig:feasibility-2}
	\end{minipage}
\end{figure}
the annual battery price reduction. Still, the realized operational costs remain the critical parameter that determines whether a sequence breaks even for \gls{abk:beb} deployment or not.

\begin{figure}[!hb]
	\centering
	\pgfplotstableread[col sep = semicolon]{content/results/baseCase/operationalDecisions/operationalDecisions_ICEB.csv}\ClusterICEB
\pgfplotstableread[col sep = semicolon]{content/results/baseCase/operationalDecisions/operationalDecisions_BEB_OC_100.csv}\ClusterOCA
\pgfplotstableread[col sep = semicolon]{content/results/baseCase/operationalDecisions/operationalDecisions_BEB_OC_200.csv}\ClusterOCB
\pgfplotstableread[col sep = semicolon]{content/results/baseCase/operationalDecisions/operationalDecisions_BEB_OC_300.csv}\ClusterOCC
\pgfplotstableread[col sep = semicolon]{content/results/baseCase/operationalDecisions/operationalDecisions_BEB_OC_400.csv}\ClusterOCD
\begin{tikzpicture}[scale=1]
	\begin{axis}[
		legend style={at={(0,1)},anchor=north west,font=\footnotesize},
		legend cell align={left},
		label style={font=\small},
		tick label style={font=\small},
		xtick pos=left,
		ytick pos=left,
		tick style = {black},
		xlabel={energy consumption [kWh]},
		ylabel={\small{max. availbale recharging time [min]}},
		ylabel near ticks,
		xmin=0,
		ymin=0,
		]
		
	\addlegendimage{mark options={mark size=2.0,fill=rwth,mark=x},only marks,draw=rwth}
	\addlegendentry{ICEB}
	
	\addlegendimage{draw=black,mark options={mark size=2.0,mark=*,fill=grun-25},only marks,draw=none,ultra thin}
	\addlegendentry{$\text{\gls{abk:beb-oc}}_{100}$}
	
	\addlegendimage{mark options={mark size=2.0,mark=triangle*,fill=grun-50},only marks,draw=none,ultra thin}
	\addlegendentry{$\text{\gls{abk:beb-oc}}_{200}$}
	
	\addlegendimage{mark options={mark size=2.0,mark=diamond*,fill=grun-75,draw=none},only marks,ultra thin}
	\addlegendentry{$\text{\gls{abk:beb-oc}}_{300}$}
	
	\addlegendimage{mark options={mark size=1.5,mark=square*,fill=grun,draw=none},only marks,ultra thin}
	\addlegendentry{$\text{\gls{abk:beb-oc}}_{400}$}

	\addplot[draw=black,mark options={mark size=2.0,mark=*,fill=grun-25},only marks,draw=none,ultra thin] table from \ClusterOCA;
	\addplot[draw=black,mark options={mark size=2.0,mark=triangle*,fill=grun-50},only marks,draw=none,ultra thin] table from \ClusterOCB;
	\addplot[draw=black,mark options={mark size=2,mark=diamond*,fill=grun-75,draw=none},only marks,ultra thin] table from \ClusterOCC;
	\addplot[draw=black,mark options={mark size=1.5,mark=square*,fill=grun,draw=none},only marks,ultra thin] table from \ClusterOCD;
	\addplot[mark options={mark size=2.0,fill=rwth,mark=x},only marks,draw=rwth] table from \ClusterICEB;
	\end{axis}
\end{tikzpicture}
	\caption{Operational assignment of bus types to trip sequences
		(\texttt{all|r\footnotesize{350}\normalsize{|b}\footnotesize{2.5}}).}
	\label{fig:operational}
\end{figure}
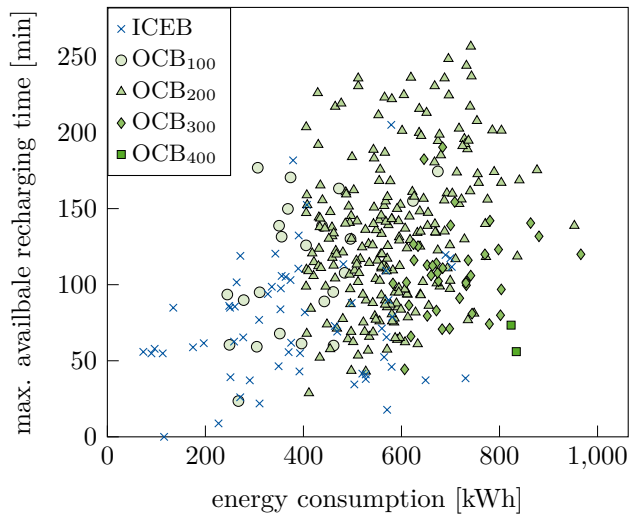

Figure~\ref{fig:operational} analyzes the assignment of bus types with respect to the characteristics of trip sequences. It details the operational assignment of buses to sequences and shows which trip sequences are operated by which bus type. Each data point represents one trip sequence~$s$ of the input vehicle schedule~$\mathcal{S}$  indicating its energy consumption and maximum available recharging time, as well as the bus type chosen to operate the sequence. Here, the maximum available recharging time is equal to the sum of available dwell-times between all trips of a sequence. As can be seen, the assignment of bus types depends on both energy consumption and potential recharging time. If feasible, \glspl{abk:beb-oc} are assigned to trip sequences with a high energy consumption to leverage the advantage of the \glspl{abk:beb}' operational cost savings. Low-capacity \glspl{abk:beb-oc} are only assigned to trip sequences with a high energy consumption if sufficient recharging time is available; in case of small available recharging times, high-capacity \glspl{abk:beb-oc} are used to operate these sequences. In contrast, \glspl{abk:iceb} are used to operate trip sequences with a low energy consumption. Thus, the missing electrification of these trip sequences is not due to range or charging limitations of \glspl{abk:beb}, but for economic reasons. 
\begin{figure}[!hb]
	\centering
	\begin{minipage}[t]{0.45\textwidth}
		\pgfplotstableread[col sep = semicolon]{content/results/baseCase/fleetComposition/fleetComposition.csv}\DataForBars 
		\begin{tikzpicture}[scale=1]
\begin{axis}[
	tick style = {black},
	label style={font=\small},
	tick label style={font=\small},
	enlarge x limits={abs=1},
	enlarge y limits=false,
	axis on top=true,
	xtick pos=left,
	ytick pos=left,
	ybar stacked,
	xtick distance=5,
	minor x tick num=4,
	xtickmin = 1,
	ymin=0,
	bar width=0.1cm,
	xlabel={period [a]},
	ylabel={number of buses [-]},
	xticklabels={,,2020,2025,2030,2035},
	]
	\addplot[fill=grun-25,ultra thin] table [x=t,y=OC100] {\DataForBars};
	\addplot[fill=grun-50,ultra thin] table [x=t,y=OC200] {\DataForBars};
	\addplot[fill=grun-75,ultra thin] table [x=t,y=OC300] {\DataForBars};
	\addplot[fill=grun,ultra thin] table [x=t,y=OC400] {\DataForBars};
	\addplot[fill=rwth,ultra thin] table [x=t,y=ICEB] {\DataForBars};
\end{axis}
\begin{axis}[
	hide axis,
	width=8.425cm,
	height=2cm,
	xmin=10,
	xmax=50,
	ymin=0,
	ymax=0.4,
	legend style={at={(0,13.625)},anchor=north west,font=\footnotesize},
	legend cell align={left},
]
	\addlegendimage{fill=rwth,mark=square*,only marks, mark size = 3.0, draw = black, ultra thin}
	\addlegendentry{\gls{abk:iceb}}
	\addlegendimage{fill=grun-25,mark=square*,only marks, mark size = 3.0, draw = black, ultra thin}
	\addlegendentry{$\text{\gls{abk:beb-oc}}_{100}$}
	\addlegendimage{fill=grun-50,mark=square*,only marks, mark size = 3.0, draw = black, ultra thin}
	\addlegendentry{$\text{\gls{abk:beb-oc}}_{200}$}
	\addlegendimage{fill=grun-75,mark=square*,only marks, mark size = 3.0, draw = black, ultra thin}
	\addlegendentry{$\text{\gls{abk:beb-oc}}_{300}$}
	\addlegendimage{fill=grun,mark=square*,only marks, mark size = 3.0, draw = black, ultra thin}
	\addlegendentry{$\text{\gls{abk:beb-oc}}_{400}$}
\end{axis}
\end{tikzpicture}
	\end{minipage}	
	\hspace{1cm}
	\begin{minipage}[t]{0.45\textwidth}		
		\vspace{-6.825cm}
		\begin{threeparttable}
			\setlength{\abovecaptionskip}{0.5ex}
			\addtolength{\tabcolsep}{-3pt}
			\small
			\def\c{\checkmark}
			\begin{tabular}{rlll}		
				\toprule
				description					& variable						& value		& unit \\
				\midrule 
				\gls{abk:tco}				& $Z$ 							& 472.22 	&\euro MM\\
				initial bus fleet value 	& $C_{t_{0}}^{\text{b}}$ 		& 54.37 	&\euro MM\\
				bus purchases				& $C_{t}^{\text{b}}$			& 209.32 	&\euro MM\\
				bus sales 					& -$V_{t}^{\text{b}}$			& -20.51	&\euro MM\\
				final bus fleet value		& -$V_{t_{\text{n}+1}}^{b}$		& -26.89 	&\euro MM\\
				charging infrastructure  	& $C_{t}^{\text{f}}$			& 2.19  	&\euro MM\\
				operational costs 			& $C_{t}^{\text{o}}$			& 253.73	&\euro MM\\
				\midrule
				\gls{abk:nox} emissions		& -								& 67.2		& t/year \\		
				\bottomrule
			\end{tabular}
			\begin{tablenotes}
				\tiny
				\item[*]
				Note that all costs are cumulated and discounted.
			\end{tablenotes}
		\end{threeparttable}
	\end{minipage}	
	\caption{Development of the bus fleet composition, \gls{abk:tco}, and \gls{abk:nox} emissions (\texttt{all|r\footnotesize{350}\normalsize{|b}\footnotesize{2.5}}).}
	\label{fig:fleet}
\end{figure}
\Glspl{abk:iceb} are used to 
operate short sequences with a low energy consumption, as for these sequences the lower operational costs of a \gls{abk:beb} cannot compensate for its higher investment costs.

\begin{figure}[!hb]
	\begin{minipage}{0.45\textwidth}
		\centering
		\begin{tikzpicture}[scale=1.175]
		\node[inner sep=0pt] at (3.75,3.05)
		{\includegraphics[width=\textwidth]{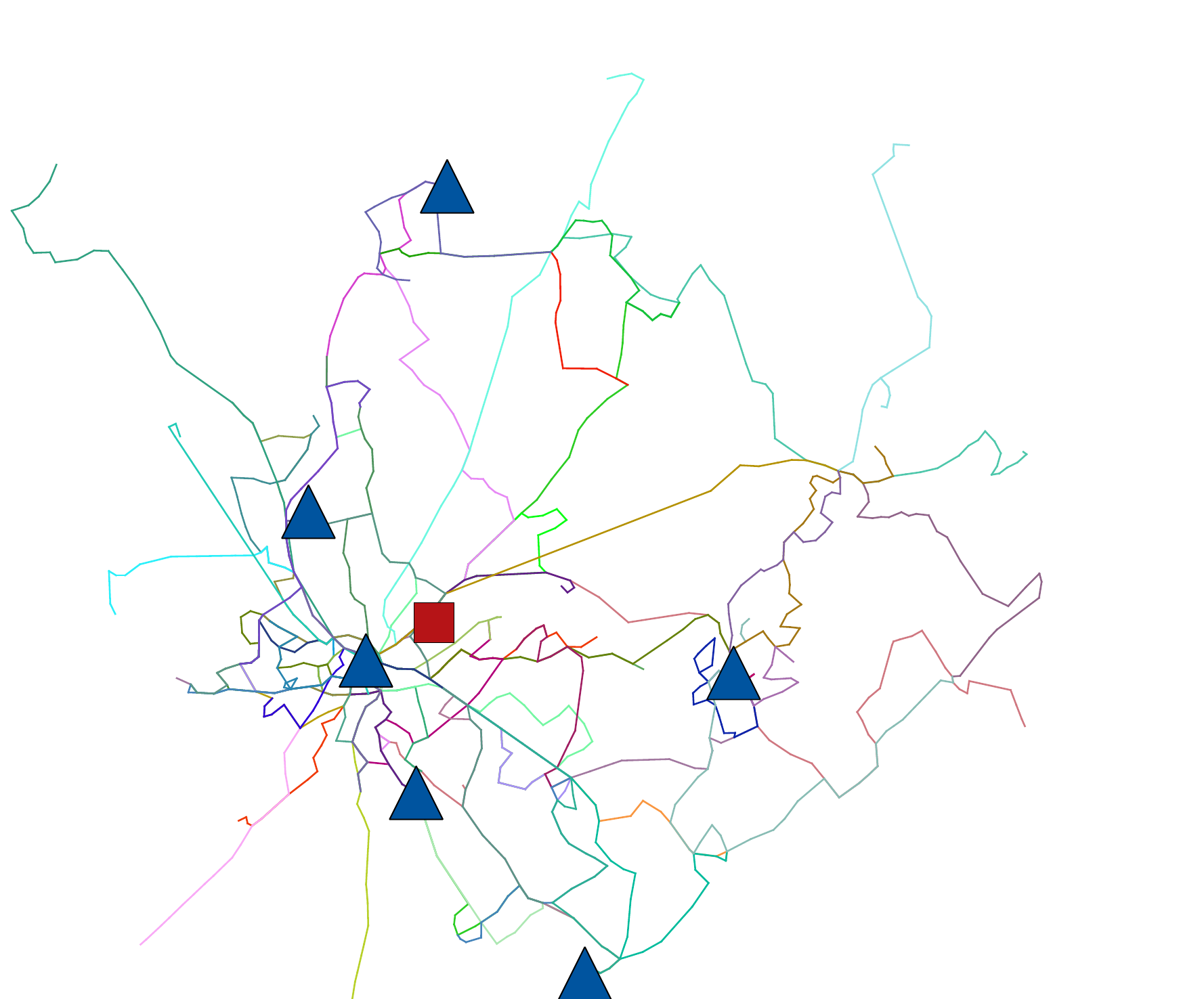}};
		\begin{axis}[%
	xmin=0,
	xmax=100,
	xticklabels={,,},
	xtick style={draw=none},
	ymin=0,
	ymax=100,
	yticklabels={,,},
	ytick style={draw=none},
	legend style={at={(1,1)},anchor=north east,font=\footnotesize},
	legend cell align={left},
	]
	\addlegendimage{rwth,only marks, mark=triangle*,draw=black,ultra thin,mark size=3.5}
	\addlegendentry{\gls{abk:frf}$\vert2020$};


	\addlegendimage{rwth, mark=none,draw=black,ultra thin}
	\addlegendentry{bus line};


	\addlegendimage{rot,only marks, mark=square*,draw=black,ultra thin,mark size=2.5}
	\addlegendentry{depot};
\end{axis}
		\end{tikzpicture}
		\caption*{(a) $t=2020$.}
	\end{minipage}
	\hspace{0.5cm}
	\begin{minipage}{0.45\textwidth}
		\begin{tikzpicture}[scale=1.175]
		\node[inner sep=0pt] at (3.75,3.05)
		{\includegraphics[width=\textwidth]{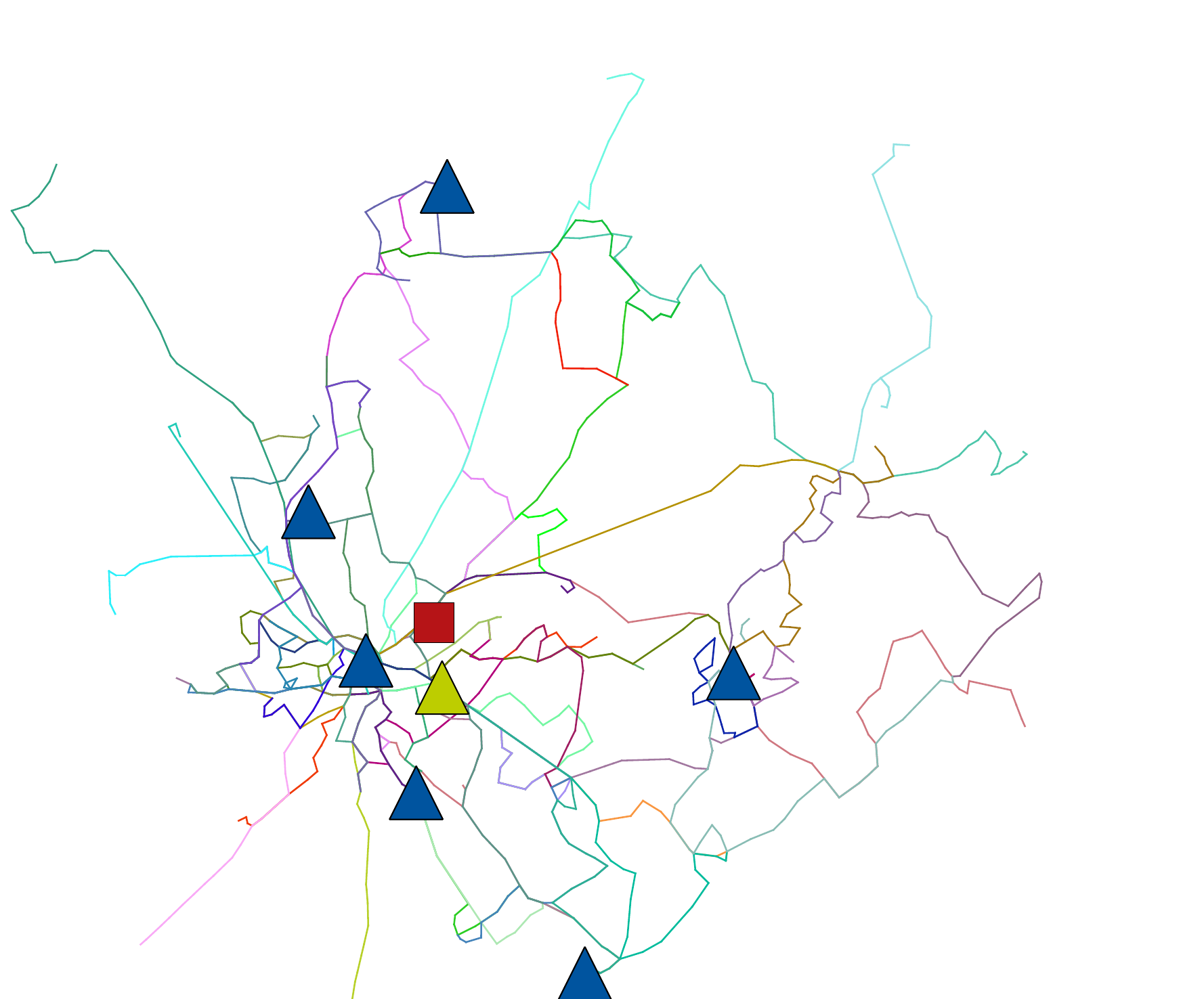}};
		\begin{axis}[%
	xmin=0,
	xmax=100,
	xticklabels={,,},
	xtick style={draw=none},
	ymin=0,
	ymax=100,
	yticklabels={,,},
	ytick style={draw=none},
	legend style={at={(1,1)},anchor=north east,font=\footnotesize},
	legend cell align={left},
	]
	\addlegendimage{rwth,only marks, mark=triangle*,draw=black,ultra thin,mark size=3.5}
	\addlegendentry{\gls{abk:frf}$\vert2020$};

	\addlegendimage{maigrun,only marks, mark=triangle*,draw=black,ultra thin,mark size=3.5}
	\addlegendentry{\gls{abk:frf}$\vert2021$};

	\addlegendimage{rwth, mark=none,draw=black,ultra thin}
	\addlegendentry{bus line};


	\addlegendimage{rot,only marks, mark=square*,draw=black,ultra thin,mark size=2.5}
	\addlegendentry{depot};
\end{axis}
		\end{tikzpicture}
		\caption*{(b) $t=2021$.}
	\end{minipage}
	\caption{Installation of opportunity charging facilities.}
	\label{fig:infrastructure}
\end{figure}

Figure~\ref{fig:infrastructure} shows the charging infrastructure investments for the optimal fleet transformation. As can be seen, all charging infrastructure investments are taken in the first and second step of the planning horizon and only few bus stations are equipped with network charging facilities. After~$2021$, no additional \glspl{abk:frf} are installed in the network. These results show that our integrated planning approach allows to identify a network design that allows to operate a large share of a bus fleet via \glspl{abk:beb} with only a few central network charging facilities. Further, these results emphasize that the electrification of trip sequences that require not only investment into \glspl{abk:beb} but also into network charging facilities takes place at the beginning of the planning horizon such that the operational cost savings can compensate the investments over the planning horizon.

\subsection{Scenario Analysis}\label{sub:scenarioAnalysis}
In the following, we summarize the findings of our sensitivity analysis for both the different bus type scenarios and for the parameter sensitivities. For detailed results, we refer to Appendix~\ref{apx:C}.

First, we discuss results on costs and \gls{abk:nox} emissions for the four bus type scenarios~\texttt{all},~\texttt{ic},~\texttt{nc}, and~\texttt{oc}. Figure~\ref{fig:tco} shows the \gls{abk:tco} structure for each bus type scenario, with all other parameters remaining as in the base case setting. Herein, information can be derived on the difference between the initial and final fleet value~($C_{0}^{\text{fleet}}-V_{\text{n}+1}^{\text{fleet}}$), investment costs resulting from the bus fleet transformation~($C_{t}^{\text{fleet}}$), operational costs~($C_{t}^{\text{oper}}$), and costs for installing charging infrastructure~($C_{t}^{\text{infr}}$). As can be seen, operational costs account for the largest share of \gls{abk:tco} in all scenarios. Scenarios~\texttt{all} and \texttt{oc} show lower operational costs but higher investment costs compared to scenarios~\texttt{ic} and \texttt{nc}. Herein, the decrease in operational costs offsets the increase in investment costs such that scenarios~\texttt{all} and~\texttt{oc} show total costs that are~$4.3$\,\% lower than in scenarios~\texttt{ic} and~\texttt{nc}. Figure~\ref{fig:emissions} shows \gls{abk:nox} emissions of all scenarios. As can be seen, scenarios \texttt{all} and \texttt{oc} result in $52.9$\,\% lower \gls{abk:nox} emissions than the \texttt{ic} and \texttt{nc} scenario.

We note that for both figures, the \texttt{all} and the \texttt{oc} as well as the \texttt{ic} and \texttt{nc} scenarios show the same results. This is not surprising because the optimal transformation strategy of the \texttt{all} scenario bases solely on the deployment of \glspl{abk:beb-oc}. Accordingly, the \texttt{nc} scenario remains equal to a pure \gls{abk:iceb} scenario.

Next, we discuss the impact of the charging power that is available for partial recharging at network charging facilities during dwell-times between trips. 
Figure~\ref{fig:results-1} shows 
\begin{figure}[!ht]
	\centering
	\hspace{-0.5cm}
	\begin{minipage}[t]{0.45\textwidth}
		\centering
		\pgfplotstableread[col sep = semicolon]{content/results/baseCase/TCO/TCO.csv}\DataForBars 
\begin{tikzpicture}[scale=1]
\begin{axis}[
tick style = {black},
label style={font=\footnotesize},
tick label style={font=\footnotesize},
enlarge x limits={abs=1},
enlarge y limits=false,
axis on top=true,
xtick pos=left,
ytick pos=left,
	legend style={at={(0.5,1)},anchor=north,legend columns=4,font=\scriptsize},
	legend style={font=\scriptsize},
	legend cell align={left},
	tick style = {black},
	label style={font=\footnotesize},
	tick label style={font=\footnotesize},	
	enlarge x limits={abs=0.75},
	xtick pos=left,
	ytick pos=left,
	xlabel near ticks,
	ylabel near ticks,
	axis on top=true,
	ybar stacked,
	bar width=1 cm,
	ymax=600,
	ymin=0,
	xtick distance=1,
	xticklabels={,,\texttt{all},\texttt{ic},\texttt{nc},\texttt{oc}},
	ylabel={TCO [\euro MM]},
	]
	\addplot[black,fill=black-10] table [x=i,y=C_b_0-V_b_n] {\DataForBars};
	\addplot[black,fill=black-50] table [x=i,y=C_b-V_b] {\DataForBars};
	\addplot[black,fill=black-75] table [x=i,y=C_v] {\DataForBars};
	\addplot[black,fill=black] table [x=i,y=C_f] {\DataForBars};
	\addlegendentry{$C_{0}^{\text{fleet}}-V_{\text{n}+1}^{\text{fleet}}$}
	\addlegendentry{$C_{t}^{\text{fleet}}$}
	\addlegendentry{$C_{t}^{\text{oper}}$}
	\addlegendentry{$C_{t}^{\text{infr}}$}
	
\end{axis}
\end{tikzpicture}
		\vspace{-0.25cm}
		\caption{\gls{abk:tco} structure for each bus type scenario with $r=350$\,kW and $b=2.5$\,\%/year.}
		\label{fig:tco}
	\end{minipage}
	\hspace{0.5cm}
	\begin{minipage}[t]{0.45\textwidth}
		\centering
		\begin{tikzpicture}[scale=1]
\begin{axis}[
	xlabel near ticks,
	ylabel near ticks,
	label style={font=\small},
	tick label style={font=\small},
	xtick distance=1,
	axis on top=true,
	xtick pos=left,
	ytick pos=left,
	tick style = {black},
	enlarge x limits={abs=0.75},
	ybar stacked,
	bar width=1 cm,
	ymin=0,
	xticklabels={,,\texttt{all},\texttt{ic},\texttt{nc},\texttt{oc}},
	ylabel={\gls{abk:nox} [t/year]},
]

\addplot[black,fill=black-50] coordinates {
	(1,67.2)
	(2,142.81)
	(3,142.81)
	(4,67.2)
};

\end{axis}
\end{tikzpicture}
		\vspace{-0.25cm}
		\caption{\gls{abk:nox} emissions for each bus type scenario with $r=350$\,kW and $b=2.5$\,\%/year.}
		\label{fig:emissions}
	\end{minipage}
\end{figure}
the \gls{abk:tco} for all bus type scenarios depending on the charging power keeping the annual battery price reduction at $2.5$\,\%/year. Obviously, the results of the \texttt{all} and \texttt{oc} scenarios as well as the results of the \texttt{ic} and \texttt{nc} scenarios are again equal as the charging power for intermediate charging does not affect the viability of \glspl{abk:beb-on}. As can be seen, the minimum \gls{abk:tco} with a reduction of $4.8$\,\% compared to the \texttt{ic} scenario can be achieved for the \texttt{all} and \texttt{oc} scenarios with a charging power of~$250$\,kW. This setting allows for the best trade-off between higher investment costs, sufficient charging power, and reduced operational costs.

Figure~\ref{fig:results-2} shows the corresponding \gls{abk:nox} emissions. As the cost savings correlate with the deployment of \glspl{abk:beb}, a maximum reduction of \gls{abk:nox} emissions by $56.5\%$ for the \texttt{all} and \texttt{oc} scenarios is also realized for a charging power of $250$\,kW. This reduction exceeds the $52.9$\,\% savings of the reference scenario, which shows that the right balance between the available charging power and its cost does not only affect a fleet's overall \gls{abk:tco} but also the viability of the electrification of single trip sequences.

We now analyse the impact of the annual battery price reduction on costs and emissions. Figure~\ref{fig:resultsBattery-1} shows the \gls{abk:tco} for all bus type scenarios depending on the annual battery price reduction while keeping the charging power at $350$\,kW. As can be seen, the deployment of \glspl{abk:beb-oc}~(\texttt{all} / \texttt{oc}) improves the \gls{abk:tco} compared to the \texttt{ic} scenario for any battery price reduction within a range of $2.2$\,\%/year to $11.6$\,\%/year. With a higher battery price reduction,
\begin{figure}[!hb]
	\centering
	\hspace{-0.5cm}
	\begin{minipage}[t]{0.45\textwidth}
		\centering
		\pgfplotstableread[col sep = semicolon]{content/results/scenarioAnalysis/recharging/recharging.csv}\DataForBars 
		\begin{tikzpicture}[scale=1]
\begin{axis}[
	legend style={at={(0,0)},font=\footnotesize,anchor=south west},
	label style={font=\footnotesize},
	tick label style={font=\footnotesize},
	ymax=500,
	ymin=430,
	legend cell align={left},
	ytick pos=left,
	ylabel={\gls{abk:tco} [\euro MM]},
	ylabel near ticks,
	xtick distance=50,
	ytick distance=10,
	tick style = {black},
	axis on top=true,
	xtick pos=left,
	xlabel={charging power level [kW]},
	label style={font=\footnotesize},
	tick label style={font=\footnotesize},
]
	
	\draw [very thick,dashed] (axis cs: 0,493.59) -- (axis cs: 550,493.59) node[pos=0.5, above] {};
	\addlegendimage{very thick,dashed}
	\addlegendentry{\texttt{ic}}

	\addplot[mark=o,black,only marks,mark size=2.5,mark options = {thick}] table [x=r,y=tco]  {\DataForBars};
	\addlegendentry{\texttt{all}}
	
	\addplot[mark=triangle,black,only marks,mark size=2.5,mark options = {thick}] table [x=r,y=depot case]  {\DataForBars};
	\addlegendentry{\texttt{nc}}
	
	\addplot[mark=diamond,black,only marks,mark size=2.5,mark options = {thick}] table [x=r,y=fast case]  {\DataForBars};
	\addlegendentry{\texttt{oc}}
\end{axis}
\end{tikzpicture}
		\vspace{-0.25cm}
		\caption{\gls{abk:tco} in scenarios \texttt{ic,all,nc,oc} for a variation of charging power (\texttt{X|r\footnotesize{50-500}\normalsize{|b}\footnotesize{2.5}}).}
		\label{fig:results-1}
	\end{minipage}
	\hspace{0.5cm}
	\begin{minipage}[t]{0.45\textwidth}
		\centering
		\pgfplotstableread[col sep = semicolon]{content/results/scenarioAnalysis/recharging/recharging.csv}\DataForBars 
		\begin{tikzpicture}[scale=1]
\begin{axis}[
	legend style={at={(0,0)},font=\footnotesize,anchor=south west},
	label style={font=\footnotesize},
	tick label style={font=\footnotesize},
	ymin=40,
	ymax=150,
	legend cell align={left},
	ytick pos=left,
	ylabel={\gls{abk:nox} [t/year]},
	ylabel near ticks,
	xtick distance=50,
	ytick distance=20,
	tick style = {black},
	axis on top=true,
	xtick pos=left,
	xlabel={charging power level [kW]},
	label style={font=\footnotesize},
	tick label style={font=\footnotesize},
	]
	\draw [very thick,dashed] (axis cs: 0,142.81) -- (axis cs: 550,142.81) node[pos=0.5, above] {};
	\addlegendimage{very thick,dashed}
	\addlegendentry{\texttt{ic}}

	\addplot[mark=o,black,only marks,mark size=2.5,mark options = {thick}] table [x=r,y=emissions]  {\DataForBars};
	\addlegendentry{\texttt{all}}
	
	\addplot[mark=triangle,black,only marks,mark size=2.5,mark options = {thick}] table [x=r,y=emissions_on]  {\DataForBars};
	\addlegendentry{\texttt{nc}}
	
	\addplot[mark=diamond,black,only marks,mark size=2.5,mark options = {thick}] table [x=r,y=emissions_oc]  {\DataForBars};
	\addlegendentry{\texttt{oc}}
	
\end{axis}
\end{tikzpicture}
		\vspace{-0.25cm}
		\caption{\gls{abk:nox} emissions in scenarios \texttt{ic,all,nc,oc} for a variation of charging power (\texttt{X|r\footnotesize{50-500}\normalsize{|b}\footnotesize{2.5}}).}
		\label{fig:results-2}
	\end{minipage}
\end{figure}
 more trip sequences break even 
\begin{figure}[!hb]
 	\centering
 	\hspace{-0.5cm}
 	\begin{minipage}[t]{0.45\textwidth}			
 		\centering
 		\pgfplotstableread[col sep = semicolon]{content/results/scenarioAnalysis/batteryPrice/batteryPrice.csv}\DataForBars 
 		\begin{tikzpicture}[scale=1]
\begin{axis}[
	legend style={at={(0,0)},font=\footnotesize,anchor=south west},
	label style={font=\small},
	tick label style={font=\small},
	ymax=500,
	ymin=430,
	legend cell align={left},
	ytick pos=left,
	ylabel={\gls{abk:tco} [\euro MM]},
	ylabel near ticks,
	xtick distance=2.5,
	ytick distance=10,
	tick style = {black},
	axis on top=true,
	xtick pos=left,
	xlabel={battery price reduction [\%/year]},
	label style={font=\small},
	tick label style={font=\small},
]
	\draw [very thick,dashed] (axis cs: -2.5,493.59) -- (axis cs: 15,493.59) node[pos=0.5, above] {};
	\addlegendimage{very thick,dashed}
	\addlegendentry{\texttt{ic}}

	\addplot[mark=o,black,only marks,mark size=2.5,mark options = {thick}] table [x=bat,y=tco]  {\DataForBars};
	\addlegendentry{\texttt{all}}
	
	\addplot[mark=triangle,only marks,mark size=2.5,mark options = {thick}] table [x=bat,y=tco_on]  {\DataForBars};
	\addlegendentry{\texttt{nc}}
	
	\addplot[mark=diamond,only marks,mark size=3,mark options = {thick}] table [x=bat,y=tco_oc]  {\DataForBars};
	\addlegendentry{\texttt{oc}}
\end{axis}
\end{tikzpicture}
 		\caption{\gls{abk:nox} emissions in scenarios \texttt{ic,all,nc,oc} for a variation of battery price reduction (\texttt{X|r\footnotesize{350}\normalsize{|b}\footnotesize{0-12.5}}).}
 		\label{fig:resultsBattery-1}
 	\end{minipage}
 	\hspace{0.5cm}
 	\begin{minipage}[t]{0.45\textwidth}	
 		\centering
 		\pgfplotstableread[col sep = semicolon]{content/results/scenarioAnalysis/batteryPrice/batteryPrice.csv}\DataForBars 
 		\begin{tikzpicture}[scale=1]
\begin{axis}[
	legend style={at={(0,0)},font=\footnotesize, anchor=south west},
	label style={font=\small},
	tick label style={font=\small},
	ymin=40,
	ymax=150,
	legend cell align={left},
	ytick pos=left,
	ylabel={\gls{abk:nox} [t/year]},
	ylabel near ticks,
	xtick distance=2.5,
	ytick distance=20,
	tick style = {black},
	axis on top=true,
	xtick pos=left,
	xlabel={battery price reduction [\%/year]},
	label style={font=\small},
	tick label style={font=\small},
]
	\draw [very thick,dashed] (axis cs: -2.5,142.81) -- (axis cs: 15,142.81) node[pos=0.5, above] {};
	\addlegendimage{very thick,dashed}
	\addlegendentry{\texttt{ic}}

	\addplot[mark=o,black,only marks,mark size=2.5,mark options = {thick}] table [x=bat,y=emissions]  {\DataForBars};
	\addlegendentry{\texttt{all}}
	
	\addplot[mark=triangle,only marks,mark size=3,mark options = {thick}] table [x=bat,y=emissions_on]  {\DataForBars};
	\addlegendentry{\texttt{nc}}
	
	\addplot[mark=diamond,only marks,mark size=3,mark options = {thick}] table [x=bat,y=emissions_oc]  {\DataForBars};
	\addlegendentry{\texttt{oc}}
\end{axis}
\end{tikzpicture}
 		\caption{\gls{abk:nox} emissions in scenarios \texttt{ic,all,nc,oc} for a variation of battery price reduction (\texttt{X|r\footnotesize{350}\normalsize{|b}\footnotesize{0-12.5}}).}
 		\label{fig:resultsBattery-2}
 	\end{minipage}
 \end{figure}
 for the \gls{abk:beb-oc} operations and thus the \gls{abk:tco} savings increase. Contrary, for \glspl{abk:beb-on}, only few trip sequences break even and the \gls{abk:tco} remain close to the \texttt{ic} scenario. This holds even for annual battery price reductions equal or higher than $7.5$\,\%/year.
 
Figure~\ref{fig:resultsBattery-2} shows the corresponding \gls{abk:nox} emissions. The emission savings for the deployment of \gls{abk:beb-oc} ranges between $42.8$\,\% and $66.5$\,\%, while the emissions of the \texttt{nc} scenarios remain close or equal to the \texttt{ic} scenario.

\subsection{Managerial Insights}\label{sub:insights}
The results of our case study allow to derive several managerial insights. We synthesize the findings of our study as follows:\\
\noindent\textbf{A comprehensive share of today's bus networks can be electrified with state-of-the-art technology:}
Our a-priori feasibility study suggests that a large share of trip sequences in existing bus networks can already be operated by \glspl{abk:beb} with available fast charging facilities and existing \gls{abk:beb} specifications. Even for the \gls{abk:aseag} network as one of the biggest European public transport systems operated solely by buses, this share amounts to $99.72$\,\%.\\
\textbf{The deployment of \glspl{abk:beb} is economically viable for a significant amount of potentially electrifiable trip sequences:}
Our results show that in the cost optimal solution, \glspl{abk:beb} are deployed on $83$\,\% out of the $99.72$\,\% of trip sequences that can potentially be electrified.\\
\textbf{Operated in mixed fleets, \glspl{abk:beb} constitute a win-win-concept:}
Our results show that, besides a significant reduction of \gls{abk:nox} emissions, an appropriate deployment of \glspl{abk:beb} in today's bus fleets may also yield economic savings with respect to a fleet's \gls{abk:tco}.\\
\textbf{(Fast) opportunity charging concepts appear to be superior to depot-charging-only concepts:}
While there is a controversial discussion whether one should \textit{i}) invest into additional fast charging infrastructure at bus stations to allow recharging during the day and decrease necessary battery capacities or \textit{ii}) compensate missing charging infrastructure by larger battery capacities and rely solely on depot charging, our results point towards a dominance of concepts that allow for additional~(partial) fast recharging in between trips.\\
\textbf{A medium-power charging infrastructure combined with medium-sized battery capacities can offset the extremes of both technology ranges:}
Our results show that instead of choosing an extreme combination of low charging power infrastructure and high battery capacities or vice versa, a balanced combination that chooses both characteristics within the middle of the two technology ranges results in a \gls{abk:tco} and NO\textsubscript{x} optimal fleet transformation.\\
\textbf{Optimal fleet transformation strategies expand over a multi-period time horizon:}
While most strategic planning approaches so far neglect a multi-period investment horizon, our results show that an optimal transformation strategy may indeed span over multiple time periods.\\ 
\textbf{Oversizing the power of charging infrastructure may decrease a fleet's cost-optimal share of \glspl{abk:beb}:} Our results show that certain trip sequences do not break even, as the potential cost savings are not sufficient to compensate the additional costs of infrastructure.

One comment on these insights is in order. By design, no case study is generic and results for other applications or case studies may differ, e.g., in case of different characteristics of the bus network or for different technology options. Accordingly, one may see our findings as a first step towards integrated (electrified) bus network design that points to interesting aspects for practitioners and show further research directions for academics. Still, we suggest to rerun the proposed methodology in practice for each application case to validate our insights.   
\section{Conclusion}\label{sect:conclusion}
We presented an optimization-based model for the cost-optimal transformation of a public transport bus network towards a~(partly) electrified fleet. Herein, we developed a \gls{abk:mip} to identify a cost-optimal, long-term, multi-period transformation plan for integrating \glspl{abk:beb} into urban bus networks while ensuring operational feasibility. We minimized the operator's \gls{abk:tco} including costs for fleet investments, charging infrastructure installation, maintenance, and operations. To this end, we considered heterogeneous fleets of \glspl{abk:iceb} and \glspl{abk:beb}, different recharging concepts, and different battery capacities. We analyzed the transformation's impact on \gls{abk:nox} emissions a-posteriori. This model is suitable for real-world applications and narrows the gap between strategic decisions and operational decisions.

We applied this modeling approach to a case study of a big European bus network in the city of Aachen, Germany, and performed additional sensitivity studies with respect to battery price reductions and charging power levels. Our results suggest that a large share of the bus network can be electrified resulting in a cost-optimal solution such that mixed fleets of \glspl{abk:beb} and \glspl{abk:iceb} constitute a win-win concept that reduces both the operator's \gls{abk:tco} and the fleet's \gls{abk:nox} emissions. Moreover, our results showed that a combination of medium-sized charging power infrastructure and medium-sized battery capacities may outperform extreme combinations of charging power and battery capacities.
 
Avenues for future research exist for a large scale impact analysis based on multiple, structurally-different case studies. In the same vein, one may consider to develop a stochastic or robust modeling approach that allows to capture uncertain energy consumption or cost developments. A methodological direction for future research is given by integrating the simultaneous computation of vehicle schedules, which are precomputed in the current modeling approach.



\singlespacing{
\bibliographystyle{model5-names}
\bibliography{literature}} 
%
\onehalfspacing
\begin{appendices}
	\normalsize
	\newpage
	\section{Schedule Computation}\label{apx:A}
Our \gls{abk:mip} in Section~\ref{sect:model} requires a vehicle schedule as input. Creating such a schedule relates to the field of vehicle scheduling problems. To determine a vehicle schedule based on a given timetable, we use a modified form of the concurrent scheduling algorithm developed by \citet{Bodin1978}, where trips, sorted by departure times, are assigned to buses under consideration of time feasibility constraints.

Figure~\ref{fig:algo} shows a pseudo-code of the modified concurrent scheduling algorithm. Given an array of trips (N), departure ($\tau^{\text{s}}$) and end ($\tau^{\text{e}}$) times of all trips, and the distance (d) and duration (t) between any pair of trips, we create a vehicle schedule (S) as follows. We sort all trips in ascending order based on departure times~(l.\,1), create an initial bus~(l.\,2) to operate the very first trip~(l.\,3), and drop this trip~(l.\,4). For all remaining sorted trips~(l.\,5), we proceed as follows. For any created bus~(l.\,6), we check if the last operated trip~(l.\,7) is a feasible predecessor of the next trip~(l.\,8). If this is the case, it stores the duration of dead-heading~(l.9). Unless no feasible bus exists~(l.\,10), we create a set of all buses with the shortest dead-heading duration~(l.\,11). Out of this set, we select all buses with the latest departure times of the last operated trip~(l.\,12). Out of this selection, the bus that so far covered the least distance~(l.\,13) is selected~(l.\,14). If there is no feasible bus at all~(l.\,15), we create a new bus~(l.\,16) and assign the next trip to this new bus~(l.\,17). Finally, we drop the assigned trip~(l.\,18).
\normalem
\begin{figure}[!hb]
	\centering
	\begin{minipage}{0.425\textwidth}
		\IncMargin{1em}
		\begin{algorithm}[H]
			\SetKwData{Left}{left}
			\SetKwData{Up}{up}
			\SetKwFunction{FindCompress}{FindCompress}
			\SetKwInOut{Input}{Input}
			\SetKwInOut{Output}{Output}
			\Indm
			\Input{N, $\tau^{\text{s}}$, $\tau^{\text{e}}$, t, d}
			\Output{S}
			\Indp
			N* $\gets$ sortAsc(N, $\tau^{\text{s}}$)\\
			B $\gets$ [0]\\
			S[0] $\gets$ N*[0]\\
			N*.pop(0)\\
			\For{j $\in$ N*}{
				\For{b $\in$ B}{
					i $\gets$ S[b][-1]\;
					\If{$\tau^{\text{e}}$[i] + t[i,j] < $\tau^{\text{s}}$[j]}{
						$\pi$[b] $\gets$ t[i,j]\\
				}}
				\eIf{$\pi$.length $>$ 0}{
					B* $\gets$ getShortest($\pi$)\\
					B** $\gets$ sortDesc(B*, S, $\tau^{\text{e}}$)\\
					B*** $\gets$ sortAsc(B**, S, d)\\
					S[B***[0]].add(j)
				}{
					B.add(B.length+1)\\
					S[B[-1]] $\gets$ [j]\\
				}
				N*.pop(0)
			}
		\end{algorithm}
		\DecMargin{1em}
	\end{minipage}
	\caption{Modified concurrent scheduling algorithm.}
	\label{fig:algo}
\end{figure}
	\clearpage
	\section{Detailed Results}\label{apx:C}
\begin{table*}[h]
	\centering
	\begin{threeparttable}
		\footnotesize
		\centering
		\setlength{\abovecaptionskip}{0.5ex}		
		\caption{Detailed results of \gls{abk:tco} and $\text{NO}_{\text{x}}$ emissions of scenarios
			\texttt{ic}
			and
			\texttt{all|r\footnotesize{0-250}\normalsize{|b}\footnotesize{0-12.5}}.}
		\label{tab:appA}
		\tabcolsep=0.1cm
		\begin{tabular}{ll|l|llll|llll|ll|ll}
			\toprule
			$r$ & $b$ &\glspl{abk:iceb}&\multicolumn{4}{c|}{\glspl{abk:beb-oc}\,[-]\tnote{*}}&\multicolumn{4}{c|}{\glspl{abk:beb-on}\,[-]\tnote{*}}& \glspl{abk:srf} & \glspl{abk:frf} & TCO & $\text{NO}_{\text{X}}$\\
			\text{[kW]}&[\%]&[-]&100&200&300&400&100&200&300&400&[-]&[-]&[\euro MM]&[t/year]\\
			\midrule
			- & - & 357 & - & - & - & - & - & - & - & - & - & - & 493.59 & 142.81\\
			50&0&294&2&29&22&10&0&0&0&0&63&0&491.02&131.49\\
50&2.5&162&2&42&59&92&0&0&0&0&201&0&486.59&118.56\\
50&5.0&137&5&47&68&100&0&0&0&0&220&0&479.8&113.73\\
50&7.5&135&7&47&68&100&0&0&0&0&223&0&473.8&111.95\\
50&10.0&134&8&47&68&100&0&0&0&0&223&0&468.83&111.26\\
50&12.5&134&8&47&68&100&0&0&0&0&223&0&464.79&110.95\\
100&0&239&4&50&60&4&0&0&0&0&118&2&487.57&119.32\\
100&2.5&100&5&65&110&77&0&0&0&0&258&2&479.84&103.02\\
100&5.0&60&8&78&120&91&0&0&0&0&297&2&470.04&97.52\\
100&7.5&58&10&78&120&91&0&0&0&0&299&2&461.76&93.93\\
100&10.0&57&11&78&120&91&0&0&0&0&300&2&455.08&84.15\\
100&12.5&58&10&78&120&91&0&0&0&0&299&0&449.92&80.39\\
150&0&206&4&100&46&1&0&0&0&0&162&7&482.95&91.41\\
150&2.5&73&5&123&115&41&0&0&0&0&284&5&473.34&75.24\\
150&5.0&29&7&140&127&54&0&0&0&0&328&3&462.49&69.58\\
150&7.5&26&10&140&127&54&0&0&0&0&332&5&453.7&62.75\\
150&10.0&25&11&140&127&54&0&0&0&0&332&2&446.28&57.81\\
150&12.5&24&12&140&127&54&0&0&0&0&333&2&440.31&56.44\\
200&0&195&7&127&28&0&0&0&0&0&184&4&480.54&85.19\\
200&2.5&56&8&171&100&22&0&0&0&0&302&4&470.54&70.27\\
200&5.0&19&9&188&110&31&0&0&0&0&338&4&459.59&62.33\\
200&7.5&15&12&189&110&31&0&0&0&0&342&4&450.32&54.24\\
200&10.0&13&14&189&110&31&0&0&0&0&344&4&442.81&53.11\\
200&12.5&12&15&189&110&31&0&0&0&0&345&5&436.98&48.66\\
250&0&179&11&147&19&1&0&0&0&0&199&5&480.08&76.88\\
250&2.5&52&13&205&72&15&0&0&0&0&305&6&469.68&62.14\\
250&5.0&15&15&222&84&21&0&0&0&0&342&5&458.27&55.56\\
250&7.5&6&18&228&84&21&0&0&0&0&351&8&449.02&50.56\\
250&10.0&4&20&228&84&21&0&0&0&0&353&6&441.09&46.71\\
250&12.5&3&21&228&84&21&0&0&0&0&354&5&434.75&43.38\\
\bottomrule

		\end{tabular}
			\begin{tablenotes}				
		\scriptsize
		\item[*] differentiated according to battery capacities (100,\,200,\,300,\,400) in kWh.
	\end{tablenotes}
	\end{threeparttable}
\end{table*}
\begin{table*}[h]
	\centering
	\begin{threeparttable}
		\footnotesize
		\centering
		\setlength{\abovecaptionskip}{0.5ex}		
		\caption{Detailed results of \gls{abk:tco} and $\text{NO}_{\text{x}}$ emissions of scenarios
		\texttt{all|r\footnotesize{300-350}\normalsize{|b}\footnotesize{0-12.5}}.}
		\label{tab:appB}
		\tabcolsep=0.1cm
		\begin{tabular}{ll|l|llll|llll|ll|ll}
			\toprule
			$r$ & $b$ &\glspl{abk:iceb}&\multicolumn{4}{c|}{\glspl{abk:beb-oc}\,[-]\tnote{*}}&\multicolumn{4}{c|}{\glspl{abk:beb-on}\,[-]\tnote{*}}& \glspl{abk:srf} & \glspl{abk:frf} & TCO & $\text{NO}_{\text{X}}$\\
			\text{[kW]}&[\%]&[-]&100&200&300&400&100&200&300&400&[-]&[-]&[\euro MM]&[t/year]\\
			\midrule
			300&0&185&16&147&9&0&0&0&0&0&196&8&481.2&80.6\\
300&2.5&54&18&223&58&4&0&0&0&0&303&7&470.61&64.92\\
300&5.0&19&20&240&69&9&0&0&0&0&338&8&459.43&56.18\\
300&7.5&8&21&249&70&9&0&0&0&0&349&7&449.7&50.31\\
300&10.0&4&23&251&70&9&0&0&0&0&353&5&441.85&48.05\\
300&12.5&3&24&251&70&9&0&0&0&0&354&6&435.66&44.81\\
350&0&194&20&140&3&0&0&0&0&0&183&9&482.81&81.73\\
350&2.5&60&24&232&39&2&0&0&0&0&297&6&472.22&67.2\\
350&5.0&19&25&251&55&7&0&0&0&0&338&7&460.83&60.58\\
350&7.5&6&27&261&56&7&0&0&0&0&351&7&451.02&53.58\\
350&10.0&3&30&261&56&7&0&0&0&0&354&8&443.14&48.57\\
350&12.5&1&31&262&56&7&0&0&0&0&356&7&436.58&47.83\\
400&0&207&25&123&2&0&0&0&0&0&164&9&484.76&89.42\\
400&2.5&66&30&234&26&1&0&0&0&0&291&9&474.57&72.7\\
400&5.0&19&31&260&41&6&0&0&0&0&338&8&462.83&64.78\\
400&7.5&5&32&272&42&6&0&0&0&0&352&7&452.77&55.45\\
400&10.0&3&34&272&42&6&0&0&0&0&354&8&444.63&50.16\\
400&12.5&1&35&273&42&6&0&0&0&0&356&7&437.94&47.0\\
450&0&231&27&98&1&0&0&0&0&0&134&8&486.44&105.2\\
450&2.5&73&34&230&20&0&0&0&0&0&284&9&476.74&85.36\\
450&5.0&19&36&270&28&4&0&0&0&0&338&9&464.89&74.14\\
450&7.5&6&38&279&29&5&0&0&0&0&351&9&454.76&62.72\\
450&10.0&3&40&280&29&5&0&0&0&0&354&8&446.3&55.85\\
450&12.5&1&42&280&29&5&0&0&0&0&356&8&439.48&50.92\\
500&0&247&30&80&0&0&0&0&0&0&114&9&487.68&111.87\\
500&2.5&90&38&213&16&0&0&0&0&0&267&9&478.85&92.47\\
500&5.0&23&41&264&26&3&0&0&0&0&334&8&467.21&83.01\\
500&7.5&7&43&275&28&4&0&0&0&0&350&9&456.84&75.1\\
500&10.0&3&46&276&28&4&0&0&0&0&354&9&448.27&64.18\\
500&12.5&1&48&276&28&4&0&0&0&0&356&9&441.26&59.41\\
\bottomrule

		\end{tabular}
		\begin{tablenotes}				
			\scriptsize
			\item[*] differentiated according to battery capacities (100,\,200,\,300,\,400) in kWh.
		\end{tablenotes}
	\end{threeparttable}
\end{table*}
\begin{table*}[h]
	\centering
	\begin{threeparttable}
		\footnotesize
		\centering
		\setlength{\abovecaptionskip}{0.5ex}		
		\caption{Detailed results of \gls{abk:tco} and $\text{NO}_{\text{x}}$ emissions of scenarios
		\texttt{nc|r\footnotesize{50-500}\normalsize{|b}\footnotesize{2.5}}.}
		\label{tab:appONrecharging}
		\tabcolsep=0.1cm
		\begin{tabular}{ll|l|llll|llll|ll|ll}
			\toprule
			$r$ & $b$ &\glspl{abk:iceb}&\multicolumn{4}{c|}{\glspl{abk:beb-oc}\,[-]\tnote{*}}&\multicolumn{4}{c|}{\glspl{abk:beb-on}\,[-]\tnote{*}}& \glspl{abk:srf} & \glspl{abk:frf} & TCO & $\text{NO}_{\text{X}}$\\
			\text{[kW]}&[\%]&[-]&100&200&300&400&100&200&300&400&[-]&[-]&[\euro MM]&[t/year]\\
			\midrule
			50&2.5&357&0&0&0&0&0&0&0&0&0&0&493.59&142.81\\
100&2.5&357&0&0&0&0&0&0&0&0&0&0&493.59&142.81\\
150&2.5&357&0&0&0&0&0&0&0&0&0&0&493.59&142.81\\
200&2.5&357&0&0&0&0&0&0&0&0&0&0&493.59&142.81\\
250&2.5&357&0&0&0&0&0&0&0&0&0&0&493.59&142.81\\
300&2.5&357&0&0&0&0&0&0&0&0&0&0&493.59&142.81\\
350&2.5&357&0&0&0&0&0&0&0&0&0&0&493.59&142.81\\
400&2.5&357&0&0&0&0&0&0&0&0&0&0&493.59&142.81\\
450&2.5&357&0&0&0&0&0&0&0&0&0&0&493.59&142.81\\
500&2.5&357&0&0&0&0&0&0&0&0&0&0&493.59&142.81\\
\bottomrule

		\end{tabular}
		\begin{tablenotes}				
			\scriptsize
			\item[*] differentiated according to battery capacities (100,\,200,\,300,\,400) in kWh.
		\end{tablenotes}
	\end{threeparttable}
\end{table*}
\begin{table*}[h]
	\centering
	\begin{threeparttable}
		\footnotesize
		\centering
		\setlength{\abovecaptionskip}{0.5ex}		
		\caption{Detailed results of \gls{abk:tco} and $\text{NO}_{\text{x}}$ emissions of scenarios
		\texttt{nc|r\footnotesize{350}\normalsize{|b}\footnotesize{0-12.5}}.}
		\label{tab:appONbatteryPrice}
		\tabcolsep=0.1cm
		\begin{tabular}{ll|l|llll|llll|ll|ll}
			\toprule
			$r$ & $b$ &\glspl{abk:iceb}&\multicolumn{4}{c|}{\glspl{abk:beb-oc}\,[-]\tnote{*}}&\multicolumn{4}{c|}{\glspl{abk:beb-on}\,[-]\tnote{*}}& \glspl{abk:srf} & \glspl{abk:frf} & TCO & $\text{NO}_{\text{X}}$\\
			\text{[kW]}&[\%]&[-]&100&200&300&400&100&200&300&400&[-]&[-]&[\euro MM]&[t/year]\\
			\midrule
			350&0&357&0&0&0&0&0&0&0&0&0&0&493.59&142.81\\
350&2.5&357&0&0&0&0&0&0&0&0&0&0&493.59&142.81\\
350&5.0&357&0&0&0&0&0&0&0&0&0&0&493.59&142.81\\
350&7.5&338&0&0&0&0&0&0&0&19&19&0&493.48&142.24\\
350&10.0&335&0&0&0&0&0&0&0&22&22&0&493.22&141.91\\
350&12.5&332&0&0&0&0&0&0&0&25&25&0&492.94&141.72\\
\bottomrule

		\end{tabular}
		\begin{tablenotes}				
			\scriptsize
			\item[*] differentiated according to battery capacities (100,\,200,\,300,\,400) in kWh.
		\end{tablenotes}
	\end{threeparttable}
\end{table*}
\begin{table*}[h]
	\centering
	\begin{threeparttable}
		\footnotesize
		\centering
		\setlength{\abovecaptionskip}{0.5ex}		
		\caption{Detailed results of \gls{abk:tco} and $\text{NO}_{\text{x}}$ emissions of scenarios
		\texttt{oc|r\footnotesize{50-500}\normalsize{|b}\footnotesize{2.5}}.}
		\label{tab:appOCrecharging}
		\tabcolsep=0.1cm
		\begin{tabular}{ll|l|llll|llll|ll|ll}
			\toprule
			$r$ & $b$ &\glspl{abk:iceb}&\multicolumn{4}{c|}{\glspl{abk:beb-oc}\,[-]\tnote{*}}&\multicolumn{4}{c|}{\glspl{abk:beb-on}\,[-]\tnote{*}}& \glspl{abk:srf} & \glspl{abk:frf} & TCO & $\text{NO}_{\text{X}}$\\
			\text{[kW]}&[\%]&[-]&100&200&300&400&100&200&300&400&[-]&[-]&[\euro MM]&[t/year]\\
			\midrule
			50&2.5&162&2&42&59&92&0&0&0&0&201&0&486.59&118.56\\
100&2.5&100&5&65&110&77&0&0&0&0&258&2&479.84&103.02\\
150&2.5&73&5&123&115&41&0&0&0&0&284&7&473.49&75.8\\
200&2.5&56&8&171&100&22&0&0&0&0&302&4&470.54&70.27\\
250&2.5&52&13&205&72&15&0&0&0&0&305&5&469.58&62.14\\
300&2.5&54&18&223&58&4&0&0&0&0&303&6&470.6&64.91\\
350&2.5&61&23&232&39&2&0&0&0&0&296&7&472.3&67.32\\
400&2.5&66&30&234&26&1&0&0&0&0&291&8&474.37&72.6\\
450&2.5&73&34&230&20&0&0&0&0&0&284&8&476.6&86.06\\
500&2.5&90&39&213&15&0&0&0&0&0&267&10&478.91&92.34\\
\bottomrule

		\end{tabular}
		\begin{tablenotes}				
			\scriptsize
			\item[*] differentiated according to battery capacities (100,\,200,\,300,\,400) in kWh.
		\end{tablenotes}
	\end{threeparttable}
\end{table*}
\begin{table*}[h]
	\centering
	\begin{threeparttable}
		\footnotesize
		\centering
		\setlength{\abovecaptionskip}{0.5ex}		
		\caption{Detailed results of \gls{abk:tco} and $\text{NO}_{\text{x}}$ emissions of scenarios
		\texttt{oc|r\footnotesize{350}\normalsize{|b}\footnotesize{0-12.5}}.}
		\label{tab:appOCbatteryPrice}
		\tabcolsep=0.1cm
		\begin{tabular}{ll|l|llll|llll|ll|ll}
			\toprule
			$r$ & $b$ &\glspl{abk:iceb}&\multicolumn{4}{c|}{\glspl{abk:beb-oc}\,[-]\tnote{*}}&\multicolumn{4}{c|}{\glspl{abk:beb-on}\,[-]\tnote{*}}& \glspl{abk:srf} & \glspl{abk:frf} & TCO & $\text{NO}_{\text{X}}$\\
			\text{[kW]}&[\%]&[-]&100&200&300&400&100&200&300&400&[-]&[-]&[\euro MM]&[t/year]\\
			\midrule 350&0&195&20&139&3&0&0&0&0&0&183&9&482.8&81.82\\
350&2.5&61&23&232&39&2&0&0&0&0&296&7&472.3&67.32\\
350&5.0&19&25&251&55&7&0&0&0&0&338&9&461.08&60.58\\
350&7.5&6&27&261&56&7&0&0&0&0&351&8&451.14&53.2\\
350&10.0&3&30&261&56&7&0&0&0&0&354&7&442.99&48.62\\
350&12.5&1&32&261&56&7&0&0&0&0&356&8&436.64&47.48\\
\bottomrule

		\end{tabular}
		\begin{tablenotes}				
			\scriptsize
			\item[*] differentiated according to battery capacities (100,\,200,\,300,\,400) in kWh.
		\end{tablenotes}
	\end{threeparttable}
\end{table*}

\end{appendices}

\end{document}